\documentclass[11pt]{article}
\usepackage[utf8]{inputenc}

\usepackage[a4paper, total={6in, 8in}]{geometry}

\usepackage{amsmath}
\usepackage{mathdots}
\usepackage{amsfonts}
\usepackage{amssymb}
\usepackage{graphicx}
\usepackage{float}
\usepackage{hyperref}
\usepackage[position=top]{subfig}
\usepackage[round, comma,sort,authoryear]{natbib}
\usepackage{booktabs}

\title{COVID-19 and Unemployment Risk: \\ Lessons for the Vaccination Campaign}
\author{Valentina Pieroni$^{1}$, Angelo Facchini$^{1*}$, Massimo Riccaboni$^{1}$  \\
        \small $^{1}$IMT School for Advanced Studies, Lucca \\
        \small $^*$ corresponding author: angelo.facchini@imtlucca.it \\
        }
\date{\today}

\begin{document}

\maketitle

\begin{abstract}
\noindent

\noindent Assessing the economic impact of COVID-19 pandemic and public health policies is essential for a rapid recovery. In this paper, we analyze the impact of mobility contraction on furloughed workers and excess deaths in Italy. We provide a link between the reduction of mobility and excess deaths, confirming that the first countrywide lockdown has been effective in curtailing the COVID-19 epidemics. Our analysis points out that a mobility contraction of 10\% leads to a mortality reduction of 5\%  whereas it leads to an increase of 50\% in full time equivalent furloughed workers. 
Based on our results, we propose a prioritizing policy for the most advanced stage of the COVID-19 vaccination campaign, considering the unemployment risk of the healthy active population.

\vspace{5mm}

\noindent {\em keywords}: COVID-19 mortality; Furlough schemes; Economic impact of lockdowns; Vaccination rollout: Unemployment risk

\end{abstract}

\section{Introduction}

The spread of the novel coronavirus (SARS-CoV-2) worldwide and the subsequent enforcement of strict containment measures by several national governments have severely impacted the world economy, which shrank by 4.3\% in 2020 \citep{Chetty2020,WB2021}.
On the supply side, social distancing measures placed workers under stay-at-home orders, shut down ‘non-essential’ activities and challenged supply chains. On the demand side, the pandemic has reduced consumer spending, virtually wiping out demand in entire economic sectors. 
A whole bunch of literature analyzes the actual effectiveness of the most restrictive policies, such as countrywide lockdown, in preventing the contagion by reducing mobility flows and discouraging social interactions \citep{acemoglu2020multi,Favero2020,yoo2020global}, pointing out the theoretical principles and frameworks behind containment policies \citep{BeselyStern2020}. In particular, recent work has shown mobility reductions to be followed by a significant drop in the number of new COVID cases and the death toll \citep{Farboodi2020,Warren2020,Glaeser2020}.

Although a stream of literature has largely investigated the epidemiological and socio-economic consequences of lockdown measures, there is still a paucity of  evidence about the effect of  reducing mobility on employment \citep{fana2020employment}. To fill this gap, in this paper, we investigate the implications of the containment policies by considering the number of working hours allowed by the Italian government to be covered by the Wage Guarantee Fund in the aftermath of the first countrywide lockdown in 2020.

This aspect is relevant for two main reasons. On the one hand, the impact of the Covid-19 crisis on Italian workers is dramatic, with 444 thousand jobs lost in 2020\footnote{As from 2020 employment statistics issued by ISTAT (Italian National Institute of Statistics) in February 2021. See \url{https://www.istat.it/it/files//2021/02/Employment-and-unemployment_202012.pdf}}. This is on top of 3.6 million furloughed workers\footnote{As from INPS data updated on October 10, 2020. For further information see \url{https://www.inps.it/nuovoportaleinps/default.aspx?itemdir=54304}}. 
On the other hand, the estimated expenditure for Covid-related Wage Guarantee Funds allowed hours is almost 20 billion Euros in 2020 \citep{EU2021}. This is by far the main Covid induced increase of public budget expenditure in Italy. This calls for an analysis of the socio-economic consequences of the vaccination roll-out strategy in Italy to speed up the recovery and to limit unemployment.  

Vaccination strategic distribution plans generally follow the WHO guidelines \citep{WHO2004} and are also consistent with the scientific literature \citep{Medlock2018, Sah2018}. Regarding Europe, recently entered in force several criteria for prioritizing population subgroups according 
to indicators like age, work, and health vulnerability. 
The Italian strategic plan, released in December 2020 \citep{Sanita2020}, provides a detailed definition of priorities involving the first administration phases covering the first nine months of 2021 \footnote{The vaccination plan has been updated following the limitation imposed to AstraZeneca vaccine and its gradual extension to people aged more than 55 years \citep{Sanita2021b}. See also \url{https://www.trovanorme.salute.gov.it/norme/renderNormsanPdf?anno=2021&codLeg=79076&parte=1\%20&serie=null}.} 
 and about 50\% of the Italian population. Regarding the last phase, specific criteria have not been provided yet. On the same page, we found other EU countries.
For instance, Germany\footnote{https://www.bundesgesundheitsministerium.de} identified six categories of prioritization: the first five categories have different urgency according to age and health risk and cover about 30 million people, whereas the sixth category includes the remaining population, covering about 45 million people. Austria\footnote{https://www.sozialministerium.at} and Switzerland\footnote{https://www.bag.admin.ch/} adopted similar rules, with a developing plan going up to the second quarter of 2021. In France, two final phases, namely 4 and 5, involve the younger population (over 18-year-old) without comorbidities, but details on allocation criteria have not been disclosed yet\footnote{https://solidarites-sante.gouv.fr/}.
On the same page is the UK plan\footnote{https://assets.publishing.service.gov.uk/}, which identifies a phase aimed at achieving coverage for the entire population, and will start after vaccinating priority groups targeting those who are at greater risk of exposure and those who provide essential public services. Ireland introduces some specifications regarding the population at lower risk\footnote{https://www.gov.ie/en/publication/39038-provisional-vaccine-allocation-groups/}, assuming that priority is given to the 18-34 age group because it includes people who have more social contacts. Among the priority groups, Ireland also identifies workers employed in essential sectors at high risk of exposure, going beyond the criterion of essentiality and looking at the riskiness in terms of exposure and contagion spread. 
The Spanish plan\footnote{ https://mscbs.gob.es} identifies some categories of the population and provides criteria for assessing their higher or lower priority. Among the categories of medium-high priority are still workers in essential sectors (to ensure the normal functioning of society) and people who are vulnerable because of their socio-economic conditions (e.g., those with precarious work, people in the lower income bracket, etc).
Indeed, drivers of socio-economic nature are found in the Irish and Spanish cases, that consider among the priority classes those who live in crowded neighborhoods or housing (therefore at high risk of an outbreak). 
There is, therefore, the need to define further criteria for the prioritization of vaccination for people not working in essential or strategic services that are substantially equivalent and generally represent a consistent share of the population. 

To this end, here we introduce a criterion for the vaccine distribution to the share of non-prioritized population, meaning the healthy and active people.    
We start providing additional evidence on the effectiveness of restriction to mobility for the Italian case. Results show that, from the public health point of view,  a ten-percent drop in mobility 
explains a 5 percent drop in excess deaths in the following month. Furthermore, we analyze the impact of mobility reduction on the Wage Guarantee Fund (number of allowed working hours) as a proxy for the suspension of the economic activity due to Covid-19 and a proxy for the induced public expenses. Results show that a 10\% drop in human mobility corresponds to a 50\% increase of the Wage Guarantee Fund (WGF) expressed in full-time equivalent units during the following month. Looking at the interpretation of a full-time equivalent unit, we can rephrase saying that a contraction in mobility explains an increase in the number of furloughed workers in the following month. 
Available data refer to a time window spanning from March to August 2020. We run a fixed-effects model on a monthly longitudinal dataset comprising 107 Italian provinces (NUTS 3 regions). For the best performance of the methods implemented, we also addressed potential endogeneity issues concerning our main variable of interest, mobility range, following an instrumental approach as in \cite{Glaeser2020}.



As a further result, we observe higher mobility associated with a larger share of essential working residents. Hence we provide evidence supporting the inclusion of workers in essential sectors among the priority categories. The intervention is even more critical with respect to those essential jobs which imply a high risk of exposure\footnote{As previously mentioned,  the Irish provisional allocation plan identifies people working in essential jobs at a high risk of exposure among the priority categories, with the rationale of minimizing harm while reducing economic disruption. A lower degree of priority is associated with workers in occupations that are essential to the functioning of society (e.g., goods-producing industries) but where preventive measures can be adopted without much difficulty. Also in Spain essential workers' priority levels are evaluated by taking into account economic criteria and by assessing the risk of exposure and of developing severe morbidity.}. Here the main aim of reducing morbidity and mortality comes together with a socio-economic rationale, as the one of reducing unemployment risk.

Concerning the advanced stage of the campaign addressing the share of the non-prioritized population, we propose to drive the allocation so that return-to-work is facilitated for 
furloughed workers, with the expected benefit of a more efficient allocation of public funds and a reduction of potential job losses. 
The criterion of prioritizing the fraction of workers with a high unemployment risk is then compared to an alternative one based on each NUTS 3 region's resident working population. Such comparison shows that the two alternatives lead in some cases to significantly different distribution priorities.

This paper is organized as follows: Section 2 provides an essential literature review about the impact of mobility restriction on socio-economic outcomes, Section 3 describes the data collection, whereas Section 4 provides a description of the econometric model. We present our results in Section 5 and discuss them in Section 6, where we also state our conclusions and future research directions. 

\section{Literature review}

The analysis of restrictive mobility policies' actual effectiveness to prevent COVID-19 infections has been addressed in a body of scientific research spanning multiple disciplines. The consequences of such policies have been examined on an international scale, and are nowadays covered by a significant and rapidly expanding literature. 
Regarding mobility restriction policies in the U.S.,  \cite{Glaeser2020} employ data on five U.S. cities to estimate the effectiveness of lockdowns and other restrictions in limiting the spread of coronavirus disease. The authors perform panel and cross-sectional regressions of the logarithm of  COVID-19 cases per capita on the percentage drop in mobility, employing the two-periods lagged value of the explanatory variable in the panel setting. To address potential endogeneity issues concerning the main regressor of interest, \textit{mobility} has been instrumented by the employment-weighted average share of essential workers\footnote{Data on essential industries from Minnesota and Delaware are used to this end.} and by the employment-weighted average telecommuting share across industries at the zip code level. According to their main instrumental variable panel specification, when controlling for ZIP code and week fixed effects, the authors find that a drop in mobility by 10 percent points leads to a 30 percent decline in COVID-19 cases per capita. In an additional specification of the cross-sectional model, they find a positive and significant relationship between the logarithm of per-capita deaths and mobility changes, which is robust to the inclusion of controls when instrumenting for mobility.  

Regarding Germany, \cite{Krenz2020} implement an instrumental variable strategy to investigate the association between COVID-19 diffusion and mobility containment at a regional level (NUTS 3 regions). As an instrument for mobility, they employ a metric assessing the quality of the road infrastructure in German regions, namely the average travel time on roads towards the next major urban center, as a proxy for remoteness. The authors argue that the impact of 'road infrastructure' on the spread of the disease goes through the effect it has on mobility flows. By regressing the logarithm of COVID-19 per-capita cases on the variation of mobility\footnote{Changes in mobility have been measured comparing mobility flows on Easter Sunday 2020 to an average Sunday in April 2019.} in an IV cross-sectional setting, this study finds a negative and significant relationship between a change in mobility and COVID-19 disease cases. According to the authors' interpretation, German regions with a higher decline in mobility on Easter Sunday have accumulated more COVID-19 cases. The IV model's first stage shows a positive relationship between mobility drops and accessibility, defined as "travel time to the next urban center", suggesting that mobility flows declined most in those areas that are less remote (i.e., metropolitan areas).  

In the Italian case,  \cite{Borsati2020} provide evidence on the association between public transport usage and the number of excess deaths, as transport modes have been addressed as a potential driver of the contagion in the ongoing debate. Using data at the local labor markets level, the authors detect a non statistically significant correlation between the propensity to use public transports and excess deaths as recorded during the first six months of 2020. They find instead a positive and significant association between the dependent variable and synthetic indices for internal and external commuting flows\footnote{Internal commuting for the local labor market (LLM) $i$ is computed as the ratio between the number of people moving between municipalities within $i$ and the population of $i$, whereas external commuting flows accounts for the number of people moving from $i$ to other LLMs and the number of people moving to $i$ from other LLMs, again normalized on LLM $i$ population.} computed on 2011 national census data, and this result is still consistent when controlling for economic and demographic variables as well as for individual and time fixed effects. 

Focusing on excess mortality, the work by \citet{Borri2020} explores the causal effect of lockdown policies in Italy on mortality by COVID-19 (again proxied by excess deaths) and mobility. Implementing a difference in differences model on a daily panel dataset, the authors show that a higher intensity of the lockdown\footnote{According to the definition given bu the authors, a municipality experiences a more intense lockdown if the reduction in the share of the active population following the lockdown is above the median reduction across all municipalities located in the same province.} 
is associated to a significant decrease in the number of excess deaths with respect to the whole population, and this holds true in particular  for older people (in the range 40-64 and beyond). A second finding is that municipalities with a higher drop in the share of active people due to the lockdown\footnote{Temporary shutdown of non essential economic activities as from DPCM March 22, 2020.} are those showing a stronger contraction in mobility.

The analysis by \cite{Bonaccorsi2020} examines the socio-economic consequences of the Italian lockdown instead. By employing a network quantity, the node efficiency, to track changes in connectivity between municipalities 14 days after the lockdown as compared to 14 days before the lockdown, the authors argue that wealthier municipalities in terms of social indicators (index of the material and social well-being) and fiscal capacity are those showing a more pronounced contraction in mobility. At the same time, however, they observe that among those municipalities experiencing a higher drop in mobility, the contraction is much higher for municipalities with a lower average income and higher levels of inequality (measured as the ratio between mean and median income). 

In this expanding stream of literature, lockdown policies have been shown to explain changes in epidemiological data often through their effects on mobility, but according \citet{Goolsbee2021},  human mobility flows (especially those accounting for consumers' visits to business locations and stores) are just partially driven by the enforcement of stay-at-home/shelter-in-place orders, as they may also arise from voluntary behavioral adjustments due to the fear of the pandemic.

Based on this literature review, we notice that although the ongoing scientific research is largely dealing with the epidemiological and socio-economic impact of the lockdown, even in terms of market labour flows \citep{Casarico2020}, there is still a lack of evidence about the effects of lockdown policies on measures which could be taken as proxies for the contraction of economic activity. 

\section{Data collection and treatment}

Data used in this paper cover the three dimensions involved in the analysis: furlough schemes, mobility, and mortality.
Furlough schemes are measured as Wage Guarantee Funds hours that have been authorized by the Italian government. Data are released by INPS (the Italian National Social Welfare Institution) and cover the period January-September 2020 \citep{INPS_CIG}. Besides, we considered the working population's share and the number of workers according to the six digits ATECO (numerical classification of economic activities, the Italian version of European NACE). Data have been collected from ORBIS database\footnote{https://www.bvdinfo.com/en-gb/our-products/data/national/aida}. We computed for each province (NUTS 3 region) the share of workers employed\footnote{We used firm-level employment data as from 2019 fiscal year reporting.} in those ATECO codes not suspended by the Italian government.

Mobility data are provided by Facebook's "Data for Good" program \citep{Maas2019}. We compared Facebook data with the census commuting data collected by the Italian statistical institute in 2011, finding that they are strongly correlated (refer to appendix \ref{app:facebook} for details). 
Finally, as representative of the epidemic spreading, we considered the excess mortality data at the municipal level collected by \citet{ISTAT_mort} expressed as the difference between the number of deaths recorded in 2020 and the average number of deaths that occurred between 2015 and 2019 in the same period. As discussed in \cite{Buonanno2020}, the excess death toll is a reliable proxy of mortality by COVID-19. Such an assumption is needed to overcome the potential issues related to the endogeneity of testing policies (especially during the first wave of the epidemics), hospital capacity, and the difference in death classification at the local level.
Table \ref{tab:data_collected} shows that data span different time and spatial resolutions, ranging from monthly data of Wage Guarantee funds to 8-hourly data of Movement Range. Regarding the spatial aggregation variability of data, we observe a variability ranging from NUTS 3 regions of  Wage Guarantee Fund to municipality level of excess mortality.

\begin{table}[t]
    \centering
     \caption{General view of the data collected}
    \begin{tabular}{lrr}
    \hline\hline
         Data & Temporal scale  & Spatial aggregation\\
         \hline
         Wage Guarantee Fund hrs. & month & NUTS3 region \\
         Excess mortality & Day & Municipality \\
         Mobility range & Day & NUTS3 region \\
         Movement between administrative regions & 8 hours  & NUTS3 region \\
         Workers by ATECO sector & Year & NUTS3 region\\
         Share of working population & Year & NUTS3 region\\
         \hline
    \end{tabular}
    \label{tab:data_collected}
\end{table}

Facebook's Data for Good program makes available different sets of data \citep{Maas2019}, covering both mobility flows between and within administrative regions. To better represent the mobility contraction inside administrative regions, we used the mobility range, an indicator that expresses the average contraction of people mobility inside an administrative region. The mobility between NUTS 3 regions has been used to compute the network centrality (or remoteness) index of Italian provinces.

As already mentioned,  we employed a measure for the number of working hours allowed to be covered with the Wage Guarantee Fund to proxy the impact of national policies and imposed shutdowns on private sector economic activities. Right after the pandemic outbreak in Italy, the Italian government has extended by decree\footnote{Decree Law n. 18/2020 issued on March 17.} the use of already existing wage guarantee schemes against the pandemic crisis to strengthen employment protection. In a joint work from INPS and Bank of Italy \citep{INPS2020}, it is reported that in March and April 2020, around $50\%$ of employers in the private sector have been allowed to use wage compensation schemes according to the new rules in force. This kind of intervention turns into lower labor costs for the firm but translates into a loss for the employee: estimates by INPS-Bank of Italy \citep{INPS2020} show a mean monthly-gross income loss of around $27\%$. Moreover, since the government grants wage subsidies, this leads to an increase in public expenditures.

Another labor market intervention in March, the firing freeze, which is a suspension of firings for the whole period we consider in our analysis, also contributed to the growing requests for wage integration schemes. Starting from around the $12^{th}$ week of 2020 (which coincides more or less with the introduction of the firing freeze and the extension of wage integration schemes), firings sharply dropped as compared to their average level in  2017-2019 \citep{Casarico2020}. Since week nine, a sharp decrease has been detected in the number of hirings as well.

National public policies have had a remarkable impact on labor market flows. According to recent estimates \citep{Viviano2020}, if measures like the extension of wage supplementation schemes, the firings freeze, and financial supports for firms had not been issued, there would have been 600 thousand more firings in 2020 because of the pandemic crisis. 

\begin{figure}[t]
\centering
\caption{Wage Guarantee Fund, excess deaths and mobility range over time}
\includegraphics[scale=0.3]{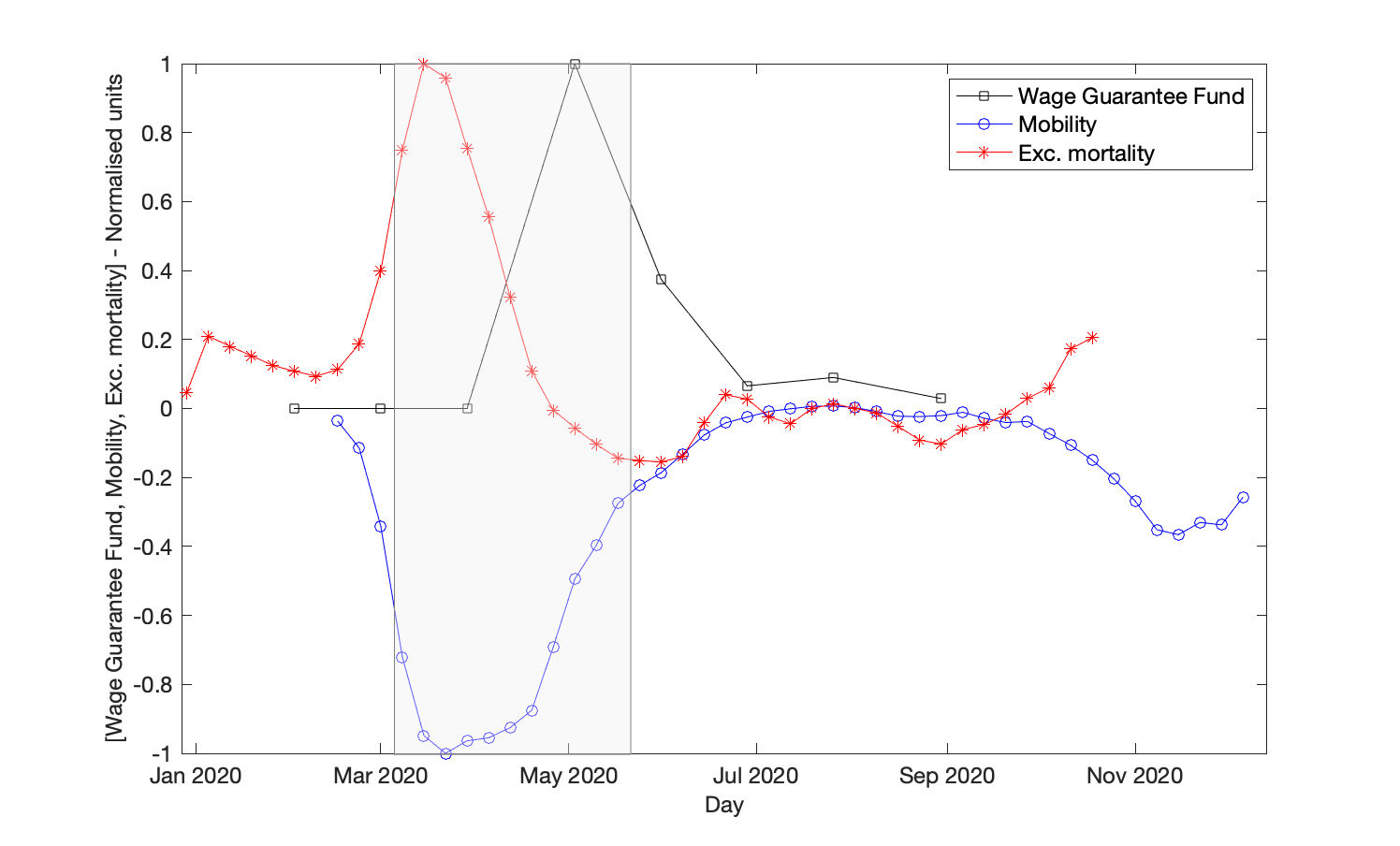}
\caption*{\footnotesize{Note: the plot displays the trend over time of the monthly average amount of the Wage Guarantee Fund (average allowed working hours), the average number of excess deaths per week and in weekly average mobility changes. All variables are expressed in normalized units: the Wage Guarantee Fund and excess deaths have been normalized on their maximum while mobility range has been normalized on its minimum. Two-weeks moving average are reported for excess deaths and mobility range.} }
\label{timeseries}
\end{figure}

Figure \ref{timeseries} shows how intense the Wage Guarantee Fund's use has been over the last year. The figure shows the monthly average Wage Guarantee Fund (in terms of accumulated hours), the weekly average number of excess deaths, and the weekly average drop in mobility evolution over a time window spanning from January to December 2020 (according to the availability of the data). Each time series has been normalized to the maximum of the period for scale uniformity and figure readability. The shaded area highlights the period when the first national lockdown was in force. As one can see, around the end of February 
mobility drops significantly (w.r.t. the baseline) while almost simultaneously the number of excess deaths shows a sharp increase reaching its peak 
around the last ten days of March. At the end of February, the first containment measures had been issued\footnote{Prime Ministerial Decree February 23, 2020.} but just on a local scale, addressing those areas where new COVID-19 cases had been recorded. However, an initial contraction of mobility flows and a growth in deaths can be detected. 
We also observe a peak in the average number of allowed working hours to be covered with the Wage Guarantee Fund at the end of April, about one month after the introduction of the national lockdown\footnote{Prime Ministerial Decree March 11, 2020} on March 12. This delay is the lag time between the working activity's temporary suspension and when the employer is allowed to use the wage guarantee schemes  \citep{INPS2020}.

While excess deaths have been computed by comparing the number of deaths in 2020 with average pre-pandemic death levels in the same time window, the amount of the Wage Guarantee Fund has not. However, the intensity in the use of the Fund in the early months of 2020 before the contagion outbreak (January and February) can be taken as a reference point. The graph shows how pre-Covid-19 levels of the number of WGF hours represent just a tiny fraction of the post-crisis levels. 

Maps in figure \ref{fig:mappeorecig}, Appendix \ref{app:maps}, instead, plot the monthly distribution of the Wage Guarantee Fund allowed hours across Italian NUTS 3 regions: darker shades point out those areas where furlough schemes (WGF hours) have been used more intensively in each month.

\begin{table}[t]
\centering
\caption{Descriptive Statistics}
\begin{tabular}{lrrrrr}
\hline\hline
 & Mean & Std. Dev. & Min. & Max. & Obs. \\ \hline
W.G.F. FTE & 14676.54 & 30937.9 & 0 & 355018.5 & 856 \\
Excess deaths & 48.884 & 264.510 & -478.2 & 5181.4 & 1070 \\
Mobility Range & -0.185 & 0.199 & -0.688 & 0.155 & 1177 \\
Betweenness & 0.030 & 0.065 & 0 & 0.581 & 855 \\
Share Essentials & 60.459 & 13.131 & 23.746 & 79.2 & 1284 \\ \hline
\end{tabular}
\label{tab:descriptives}
\end{table}


Before composing the panel, data have been checked for consistency. They have been averaged/rescaled whenever possible to fit the weekly variation and the spatial aggregation of a NUTS 3 administrative region.
As a result, we obtained a longitudinal dataset comprising monthly observations on a cross-section of 107 Italian NUTS 3 regions. 

\section{The econometric model}

To explore the relationship between the dependent variables and the main explanatory variable, namely {\itshape Mobility Range},  we employed a linear model for longitudinal data, accounting for Italian NUTS 3 regions' unobserved heterogeneity. 
The linear model is expressed as follows

\begin{equation}
ln(y)_{it}=\beta Mob.Range_{i(t-1)} + \delta Lockdown_{t} + pv_i + \varepsilon_{it}
\label{FEmodel}
\end{equation}


\noindent where $pv_i$ denotes the individual-specific fixed effects, controlling for NUTS 3 regions' unobserved time-invariant characteristics. Since $pv_i$ might be correlated with the observed regressors, we implemented a fixed-effects model.
The model also includes a dummy named {\itshape Lockdown} which takes value 1 in those months when the national lockdown was in force - March, April, and May - to control for potential time-varying effects due to the imposed restrictions. 

We test two versions of the model where $y_{it}$ of equation (\ref{FEmodel}) stands for {\itshape Excess Deaths} or {\itshape Wage Guarantee Fund}, respectively. In both cases the logarithm of the dependent variable has been regressed on a month lagged value of the explanatory variable {\itshape Mobility Range}. 
Usually, a delay of about a month occurs between the time in which a firm requires the wage supplementation schemes and the time it is officially authorized and recorded  \citep{INPS2020}. Similarly, changes in the mobility induced by restriction measures are followed with some delay by a decrease of COVID-19 deaths \citep{Borri2020}.

We use instrumental variables (IVs) to overcome potential endogeneity issues concerning our main explanatory variable {\itshape Mobility Range}. 

As already pointed out in previous scientific works, mobility is likely to be endogenous with variables strictly related to the spread of the disease, such as the number of COVID-19 cases or excess deaths \citep{Glaeser2020,Krenz2020,Borri2020}. A potential reversed causality issue may affect the estimates since mobility flows are adjusted when people observe an increase (or decrease) in the spread of COVID-19.

A similar argument applies to the relationship between mobility and the amount of the Wage Guarantee Fund. Reduced mobility can explain an increase in the Wage Guarantee Fund since the enforcement of containment measures meant to discourage mobility and limit social interactions could foster the use of wage guarantee schemes to reduce physical proximity in the workplace.
However, temporary suspension of working activities could itself explain a further drop in commuting flows. This could be the most intuitive way to interpret the relationship between mobility and furlough schemes but is not the only one, as mobility could impact the Wage Guarantee Fund even through different channels. 
If fewer people move because of containment rules or fear of contagion, we may observe a decline in the demand for goods and services by final consumers. In turn, employers may opt for a temporary reduction of working time and ask for wage compensation schemes to cope with a contraction in the demand\footnote{Even as a potential effect of consumer substitution patterns \citep{Goolsbee2021}.}.

To overcome potential endogeneity-issues, we employed and IV strategy for {\itshape Mobility Range}, testing different specifications of our models. 

The choice of the first instrument is inspired \footnote{We refer to  \cite{Glaeser2020} in the choice of the instrument for mobility, but we computed the measure according to a different formula better suited for the Italian case.} by \citet{Glaeser2020}. Looking at the provisions of Prime Ministerial Decrees issued between March and May\footnote{Our references are Dpcm March 11, Dpcm March 22, Dpcm April 1, Dpcm April 10, Dpcm April 26 and Dpcm May 17, 2020.} we computed 
the time-varying share of essential residents ({\itshape Share Essentials}) in each NUTS 3 region, that is, the share of labor force which was allowed to move during the first national lockdown since employed in economic sectors designated as essential by the Italian government. The share of authorized employees has been multiplied by the 2019 employment rate of Italian provinces \footnote{Source of employment data is ISTAT (Italian National Institute of Statistics) Labour Force Survey.} to proxy the share of essential workers of NUTS 3 regions.    

As a second instrumental variable (IV), we computed the time-varying centrality of NUT3 regions in the national mobility flows. We compute the betweenness centrality \citep{Newman} of Italian NUTS 3 regions  in the mobility network built on Facebook data movement between administrative regions\footnote{As alternative measures of centrality we computed also the Pagerank, and the variation in nodal efficiency as in \citet{Bonaccorsi2020}. We performed several trials employing each quantity alone and combined as IVs in the econometric model. We finally opted not to use more than one network quantity as an instrument (e.g. when the page rank is used as an excluded instrument together with the betweenness it appears to be redundant), and we chose the betweenness to be used alone and combined with the share of essential residents.}. A low value of the betweenness centrality is a proxy of Italian NUTS 3 regions' remoteness, in the same vein as in \citet{Krenz2020}.\footnote{The  betweenness centrality provides different information with respect to the one conveyed by mobility range. It describes a global property of the network connecting Italian provinces (NUTS 3 regions).} Mobility range measures the average reduction of mobility within an administrative region. In contrast, the betweenness centrality looks at the whole network, in a global perspective, providing a ranking of Italian NUTS 3 areas based on their importance of bridging different regional mobility systems. Based on this, we assume this quantity to be less or even not susceptible to changes in the number of fatalities or the number of Wage Guarantee Fund allowed hours at a local scale (i.e. changes referring to the single NUTS 3 region).

 Following the argument in \citet{Glaeser2020} and \citet{Krenz2020}, we assume that the centrality of a territorial unit in the mobility network and the share of people employed in essential industries have an impact on excess deaths just through mobility flows. A similar argument applies for Wage Guarantee Funds allowed hours. 

The first stage and main equations for the IV model are given by

\begin{eqnarray}
Mob.Range_{i(t-1)} & = & \pi IV_{i(t-1)} + \gamma Lockdown_{t} + pv_i + \eta_{it} \\
 ln(y)_{it} & = &   \beta Mob.Range_{i(t-1)} + \delta Lockdown_{t} + pv_i + \varepsilon_{it} 
 \label{IVmodel}
\end{eqnarray} 

\noindent both stages control for NUTS 3 region-specific fixed effects and include the {\itshape Lockdown} dummy.

We perform a GMM estimation of the coefficients of the model. Specifically, we estimate three specifications of the model: the first one employs only betweenness centrality as an IV, the second one includes just {\itshape Share Essentials}, the third model uses both variables as instruments. As a robustness check, in the case of Wage Guarantee Funds, we use local rainfalls as an alternative IV for mobility. To be aligned with the instrumented variable, all IVs have been lagged by one month. 

\section{Results} 
 
Our analysis estimates the impact of mobility range, which keeps track of the change in mobility that occurred in Italy during and after the first national lockdown on the logarithm of excess deaths (see table \ref{mortfullsplit}).\footnote{To cope with negative values, we first rescaled excess deaths by adding the absolute value of its minimum (i.e., $478.2$) plus one to each observation, then we took the logarithm.}

Model [A] considers a panel comprising all 107 Italian NUTS 3 regions and a time window spanning from March 2020 to October 2020.
Column (1) reports fixed-effects estimates from model \eqref{FEmodel}
while columns from (2) to (4) display the results stemming from the instrumental variable approach. 
To overcome potential endogeneity issues concerning our main regressor, in column (2), we instrument for mobility range with the betweenness centrality. 
The specification in column (3) employs share essentials as the external instrument, and the model in column (4) uses both time-varying IVs. 

As previously mentioned, each specification includes the {\itshape Lockdown} dummy variable, which takes value 1 in those months when the first national lockdown was in force in Italy:  March, April, and May. 

All models report a positive and statistically significant impact of mobility of excess deaths in the full-length period. 

Model (1) shows that excess deaths increase by around 0.3 percent at time $t$ if mobility increases by one percent in the previous month.\footnote{About the interpretation of the coefficient: since we are dealing with semi-elasticities we say that a unit increase in mobility implies a $(\beta*100$ )\% variation in the dependent variable. 'Mobility range' is not expressed in percentage points, meaning that a unit change in mobility actually means a 100$\%$ change (in order to be expressed in percentage point it should be multiplied by 100).}
Point estimates from specification (1) should be interpreted with care. 
Indeed, the magnitude of the effect grows as we instrument our main regressor, as in specifications (2)-(4), suggesting a downward bias potentially due to endogeneity issues.

Model estimates have been repeated by splitting the sample into two periods, one comprising months from March to May (model [B]) when the national lockdown was in force ({\itshape lockdown} period), and the other including observations from June to October (model [C]). 
As expected, mobility led to more deaths during the lockdown period (model [B]), when the first wave of the pandemic reached its peak in Italy, and this effect weakened in the following months (model [C]). All coefficients are significant except for model [C](2), the one instrumenting for NUTS 3 regions' centrality in the post-lockdown period.

Furthermore, results in section [B] display an increase in the coefficient once we instrument the main explanatory variable. In contrast, IV estimates in section [C] tend to be lower than in the baseline model [C](1) when statistically significant. 

 Model [A] in table \ref{mortfirstred} displays results from the first stage of the IV model. The share of essential residents shows a positive and significant relationship with mobility, meaning that, as expected, an increase in the fraction of working people employed in essential activities implies a lower drop in mobility. Concerning the {\itshape Betweenness} centrality, we observe a negative and significant coefficient instead, suggesting a higher contraction in mobility flows for NUTS 3 regions with a higher centrality in the national network. 

All in all, we find a strong and statistically significant effect of mobility on excess deaths in Italy, with a much stronger impact during the nationwide lockdown phase.

In the second part of our analysis, we estimate the potential negative consequences of reducing mobility flows on employment, as measured by the Wage Guarantee Fund. 
 
As for excess mortality, the model has been estimated on the whole period between March to August (model version [A]) and for two phases when the national lockdown was in force  (model [B] - March to May) and in the following months (model [C] - June to August).
 
Results are reported in table \ref{CIGfullsplit}. The full sample regression shows a negative and significant effect of  {\itshape Mobility Range} on furloughed workers. For every one percent drop in mobility at time $(t-1)$, we observe a $3.25$ percent increase in the use of Wage Guarantee Fund in the following month, according to model (2) with the betweenness centrality as an IV.\footnote{Again, recall that the explanatory variable Mobility Range is not expressed in percentage units and should be multiplied by 100 to be so.} 

 Our results highlight that the enforcement of national policies which have been effective in curtailing the contagion discouraged mobility flows and fostered the use of wage compensation schemes provided by law\footnote{Decree Law No. 18/2020 of 17 March 2020} to support workers. 

Estimates of the effect of mobility on wage compensation schemes tend to decrease when we instrument our main regressor by just one or both the selected IVs (column (2)-(4)). However, results in column (4) should be taken carefully since we reject the null hypothesis from the Sargan-Hansen test of over-identifying restrictions (Hansen J statistic=$9.961$). 
 
Model [B] from table \ref{CIGfullsplit} confirms our expectations: the impact of movement range on wage compensation schemes is stronger during the lockdown period when stricter restrictions were in force. Point estimates in model [C] show a sharp drop in magnitude instead: after a gradual easing of containment measures between June and August, we still see a negative association between changes in mobility and the authorized Wage Guarantee Fund but this relationship seems to be just slightly significant according to the IV specification in column (3) (t-statistic$=-1.67$,  p-value$= 0.097$) or no significant at all as from columns (2)  (t-statistic$=-0.31$,  p-value$=0.760$) and (4) (t-statistic$=-1.64$,  p-value$=0.102$).\footnote{When focusing on the sub-periods (model [B] and [C]) and instrumenting by both the IVs (column (4)) we do not reject the null hypothesis from the Sargan-Hansen test.}
 
First stage regressions results (section [A], table \ref{CIGfirstred}) are in agreement with what already displayed in table \ref{mortfirstred}, that is, a higher drop in mobility is observed in more central NUTS 3 regions and in those showing a lower fraction of essential working residents. 
 
To corroborate our results on the association between mobility and furlough schemes, we performed a robustness check exploiting the exogenous variation in rainfall as an instrument for mobility. The validity of this IV relies on the exogenous nature of weather shocks and on the assumption that an induced change in mobility is the main channel through which weather conditions (rainfall specifically) may affect the use of furlough schemes.\footnote{Rainfall shocks have been widely used in the econometric literature as instruments 
\citep{Bohlken2010,Miguel2004,SandholtJensen2009,Sarsons2015,muscillo2021}.}   
 
Table \ref{reg:cig_rainfall} (appendix \ref{app:robustness}) displays the full sample and split-sample estimates when instrumenting for mobility by the share of rainy days recorded in a month, namely {\itshape Rainfall share}. We still find a negative and statistically significant relationship between mobility and furlough schemes, consolidating our previous results. Once again, we observe a more intense impact of mobility shocks on furlough schemes during the lockdown period; in the post-lockdown period, mobility explains a lower change in the Wage Guarantee Fund, but the estimates do not lose statistical significance. 
 
 In the first stage regression, rainfall has an expected negative and statistically significant effect on mobility.\footnote{More details about the rationale behind this exercise are provided in appendix \ref{app:robustness}.}   
 
Maps in figure \ref{mapsFE}  provide a graphical representation of the NUTS 3 region-specific fixed effects estimates obtained from the main equation of the instrumental variable model (equation \ref{IVmodel}) when instrumenting by centrality and share essentials (tables \ref{mortfullsplit} and \ref{CIGfullsplit}, section [A], column (4)). 
 
 

The intensity of the color filling each territorial unit on the map is proportional to the associated individual fixed-effect coefficient: darker colors express higher coefficients. NUTS 3 regions whose coefficients are not statistically significant are in grey. The baseline in both cases (panel 'a' and 'b') is the Province of Agrigento.
 
 Concerning excess deaths, Figure \ref{mapsFE}(above) shows that the NUTS 3 regions which explain a higher increase in the dependent variable in the whole period (once controlling by mobility and lockdown-related time trends) are the most affected provinces located in the north of Italy. This somehow reflects the uneven spread of the disease across Italian regions \citep{Ascani2020}.

The map is less polarized in the case of the Wage Guarantee Fund (see Figure \ref{mapsFE}(below)). However, we still find higher coefficients in the north-east of the country plus some darker areas located in the north-west and central regions.
 
Following the same approach, estimates for the NUTS 3 region-specific fixed effects in the lockdown and post-lockdown periods only (models [B] and [C] in tables \ref{mortfullsplit} and \ref{CIGfullsplit}, specification (4)) are displayed in Figure \ref{fig:felockpost}, Appendix \ref{app:maps}. While fixed effects coefficients seem to be heterogeneous and geographically clustered when the lockdown was in force, we observe more homogeneous effects as the lockdown measures have been loosened.
 
\begin{table}[H]
\centering
\def\sym#1{\ifmmode^{#1}\else\(^{#1}\)\fi}
\caption{Excess deaths panel results}
\resizebox*{!}{\textheight}{ 
\begin{tabular}{l*{4}{c}}
\\
 \\  
 
 &\multicolumn{4}{c}{$\mbox{ln Excess Deaths}_{it}$}\\
 \cline{2-5}
 \\
                   &\multicolumn{4}{c}{{\bfseries [A] Full Sample Regression}}\\\\ &\multicolumn{1}{c}{(1)}&\multicolumn{1}{c}{(2)}&\multicolumn{1}{c}{(3)}&\multicolumn{1}{c}{(4)}\\
                    &\multicolumn{1}{c}{FE}&\multicolumn{1}{c}{IV}&\multicolumn{1}{c}{IV}&\multicolumn{1}{c}{IV}\\
\cline{2-5}
$\mbox{Mobility range}_{i(t-1)}$&       0.321\sym{***}&       0.733\sym{***}&       0.492\sym{***}&       0.491\sym{***}\\
                    &     (0.042)         &     (0.201)         &     (0.060)         &     (0.061)         \\
[1em]
Lockdown            &       0.252\sym{***}&       0.374\sym{***}&       0.303\sym{***}&       0.300\sym{***}\\
                    &     (0.038)         &     (0.067)         &     (0.032)         &     (0.032)         \\
[1em]
Constant            &       6.231\sym{***}&                     &                     &                     \\
                    &     (0.007)         &                     &                     &                     \\
\hline
Observations        &         856         &         855         &         856         &         855         \\
Number Ids                 &     107         &     107         &     107       &     107       \\
Individual FE           & Yes &   Yes  &  Yes &        Yes                        \\
Overall $R^2$                &       0.137         &                     &                     &                     \\
Root MSE            &       0.188         &       0.215         &       0.204         &       0.204         \\
First Stage F-Stat.                 &  &  53.64 &     565.13       &              292.32   \\
Hansen J stat.                   &                     &                     &                     &       1.550         \\


\toprule

 \\  
 &\multicolumn{4}{c}{{\bfseries [B] Split sample regression (March to May)}}\\\\
                    &\multicolumn{1}{c}{(1)}&\multicolumn{1}{c}{(2)}&\multicolumn{1}{c}{(3)}&\multicolumn{1}{c}{(4)}\\
                    &\multicolumn{1}{c}{FE}&\multicolumn{1}{c}{IV}&\multicolumn{1}{c}{IV}&\multicolumn{1}{c}{IV}\\
\cline{2-5}
$\mbox{Mobility range}_{i(t-1)}$&       0.367\sym{***}&       1.095\sym{***}&       0.620\sym{***}&       0.594\sym{***}\\
                    &     (0.057)         &     (0.355)         &     (0.077)         &     (0.074)         \\
[1em]
Constant            &       6.500\sym{***}&                     &                     &                     \\
                    &     (0.021)         &                     &                     &                     \\
\hline
Observations        &         321         &         320         &         321         &         320         \\
Number Ids                 &     107        &     107         &     107        &     107        \\
Individual FE           & Yes &   Yes  &  Yes &        Yes                        \\
Overall $R^2$                &       0.040         &                     &                     &                     \\
Root MSE            &       0.190         &       0.317         &       0.244         &       0.242         \\
First Stage F-Stat.                   &   &   23.25         &            769.74  &   379.77 \\
Hansen J stat.                   &                     &                     &                     &       2.104         \\


\toprule

 \\  
                   &\multicolumn{4}{c}{{\bfseries [C] Split sample regression (June to October)}}\\\\
&\multicolumn{1}{c}{(1)}&\multicolumn{1}{c}{(2)}&\multicolumn{1}{c}{(3)}&\multicolumn{1}{c}{(4)}\\
                    &\multicolumn{1}{c}{FE}&\multicolumn{1}{c}{IV}&\multicolumn{1}{c}{IV}&\multicolumn{1}{c}{IV}\\
\cline{2-5}
$\mbox{Mobility range}_{i(t-1)}$&       0.219\sym{***}&       0.309         &       0.197\sym{***}&       0.193\sym{***}\\
                    &     (0.033)         &     (0.245)         &     (0.037)         &     (0.036)         \\
[1em]
Constant            &       6.224\sym{***}&                     &                     &                     \\
                    &     (0.002)         &                     &                     &                     \\
\hline
Observations        &         535         &         535         &         535         &         535         \\
Number Ids                &     107        &     107         &     107        &     107       \\
Individual FE           & Yes &   Yes  &  Yes &        Yes                        \\
Overall $R^2$                &       0.053         &                     &                     &                     \\
Root MSE            &       0.085         &       0.096         &       0.095         &       0.095         \\
First Stage F-Stat.                   &   &     27.80       &            327.83  & 191.27  \\
Hansen J stat.                   &                     &                     &                     &       0.220         \\
\hline\hline
\multicolumn{5}{l}{\footnotesize Robust standard errors in parentheses}\\
\multicolumn{5}{l}{\footnotesize \sym{*} \(p<0.10\), \sym{**} \(p<0.05\), \sym{***} \(p<0.01\)}\\
\end{tabular}}
\label{mortfullsplit}
\end{table}

\begin{table}[H]
\centering
\def\sym#1{\ifmmode^{#1}\else\(^{#1}\)\fi}
\caption{Excess deaths IV panel results}
\resizebox{11cm}{!}{
\begin{tabular}{l*{3}{c}}

\\ 
&\multicolumn{3}{c}{{\bfseries [A] Full Sample First Stage IV}}\\\\
 &\multicolumn{1}{c}{(2)}&\multicolumn{1}{c}{(3)}&\multicolumn{1}{c}{(4)}\\
                
                    &\multicolumn{1}{c}{$\mbox{Mobility}$}&\multicolumn{1}{c}{$\mbox{Mobility}$}&\multicolumn{1}{c}{$\mbox{Mobility}$}\\
                    &\multicolumn{1}{c}{$\mbox{range}_{i(t-1)}$}&\multicolumn{1}{c}{$\mbox{range}_{i(t-1)}$}&\multicolumn{1}{c}{$\mbox{range}_{i(t-1)}$}\\
\cline{2-4}
$\mbox{Betweenness}_{i(t-1)}$&      -1.376\sym{***}&                     &      -0.409\sym{***}\\
                    &     (0.188)         &                     &     (0.150)         \\
[1em]
Lockdown            &      -0.286\sym{***}&      -0.102\sym{***}&      -0.102\sym{***}\\
                    &     (0.015)         &     (0.012)         &     (0.012)         \\
[1em]
$\mbox{Share essentials}_{i(t-1)}$&                     &       0.017\sym{***}&       0.017\sym{***}\\
                    &                     &     (0.001)         &     (0.001)         \\
\hline
Observations        &         855         &         856         &         855         \\
Number Ids                 &     107         &     107        &     107         \\
Root MSE               &       0.180         &       0.120         &       0.119         \\
Individual FE                &                    Yes &        Yes             &          Yes           \\
\toprule

\\ 
&\multicolumn{3}{c}{{\bfseries [B] Full Sample Reduced form IV}}\\\\
&\multicolumn{1}{c}{(2)}&\multicolumn{1}{c}{(3)}&\multicolumn{1}{c}{(4)}\\

                    &\multicolumn{1}{c}{$\mbox{ln Excess}$}&\multicolumn{1}{c}{$\mbox{ln Excess}$}&\multicolumn{1}{c}{$\mbox{ln Excess}$}\\
                    &\multicolumn{1}{c}{$\mbox{Deaths}_{it}$}&\multicolumn{1}{c}{$\mbox{Deaths}_{it}$}&\multicolumn{1}{c}{$\mbox{Deaths}_{it}$}\\
\cline{2-4}
$\mbox{Betweenness}_{i(t-1)}$&      -1.008\sym{***}&                     &      -0.547\sym{**} \\
                    &     (0.283)         &                     &     (0.265)         \\
[1em]
Lockdown            &       0.164\sym{***}&       0.253\sym{***}&       0.252\sym{***}\\
                    &     (0.018)         &     (0.025)         &     (0.025)         \\
[1em]
$\mbox{Share essentials}_{i(t-1)}$&                     &       0.009\sym{***}&       0.008\sym{***}\\
                    &                     &     (0.001)         &     (0.001)         \\
\hline
Observations        &         855         &         856         &         855         \\
Number Ids                 &     107         &     107         &     107         \\
Root MSE                &       0.208         &       0.198         &       0.197         \\
Individual FE                &                    Yes &        Yes             &          Yes           \\
\hline\hline
\multicolumn{4}{l}{\footnotesize Robust standard errors in parentheses}\\
\multicolumn{4}{l}{\footnotesize \sym{*} \(p<0.10\), \sym{**} \(p<0.05\), \sym{***} \(p<0.01\)}\\
\end{tabular}}
\label{mortfirstred}
\end{table}

\begin{table}[H]
\centering
\def\sym#1{\ifmmode^{#1}\else\(^{#1}\)\fi}
\caption{Wage Guarantee Fund panel results}
\resizebox*{!}{\textheight}{
\begin{tabular}{l*{4}{c}}
\\
 &\multicolumn{4}{c}{$\mbox{ln Wage Guarantee Fund FTE}_{it}$}\\\\
 \cline{2-5}
 \\
                   &\multicolumn{4}{c}{{\bfseries [A] Full Sample Regression}}\\\\ &\multicolumn{1}{c}{(1)}&\multicolumn{1}{c}{(2)}&\multicolumn{1}{c}{(3)}&\multicolumn{1}{c}{(4)}\\
                    &\multicolumn{1}{c}{FE}&\multicolumn{1}{c}{IV}&\multicolumn{1}{c}{IV}&\multicolumn{1}{c}{IV}\\
\cline{2-5}
$\mbox{Mobility range}_{i(t-1)}$&      -5.680\sym{***}&      -3.254\sym{***}&      -5.304\sym{***}&      -5.218\sym{***}\\
                    &     (0.192)         &     (0.748)         &     (0.242)         &     (0.242)         \\
[1em]
Lockdown            &      -1.673\sym{***}&      -1.051\sym{***}&      -1.576\sym{***}&      -1.564\sym{***}\\
                    &     (0.093)         &     (0.234)         &     (0.125)         &     (0.126)         \\
[1em]
Constant            &       8.273\sym{***}&                     &                     &                     \\
                    &     (0.019)         &                     &                     &                     \\
\hline
Observations        &         619         &         618         &         619         &         618         \\
Number Ids                &     104       &     104         &     104   &     104       \\
Individual FE           & Yes &   Yes  &  Yes &        Yes                        \\
Overall $R^2$              &       0.438         &                     &                     &                     \\
Root MSE            &       0.890         &       1.099         &       0.978         &       0.981         \\
First Stage F-Stat.                 &  & 49.29  &     599.79      &  305.92             \\
Hansen J stat.                   &                     &                     &                     &       9.961         \\
\toprule


 \\
                   &\multicolumn{4}{c}{{\bfseries [B] Split sample regression (March to May)}}\\\\

&\multicolumn{1}{c}{(1)}&\multicolumn{1}{c}{(2)}&\multicolumn{1}{c}{(3)}&\multicolumn{1}{c}{(4)}\\
                    &\multicolumn{1}{c}{FE}&\multicolumn{1}{c}{IV}&\multicolumn{1}{c}{IV}&\multicolumn{1}{c}{IV}\\
\cline{2-5}
$\mbox{Mobility range}_{i(t-1)}$&      -7.089\sym{***}&      -4.831\sym{***}&      -5.585\sym{***}&      -5.573\sym{***}\\
                    &     (0.236)         &     (1.079)         &     (0.263)         &     (0.264)         \\
[1em]
Constant            &       6.070\sym{***}&                     &                     &                     \\
                    &     (0.089)         &                     &                     &                     \\
\hline
Observations        &         307         &         306         &         307         &         306         \\
Number Ids                &     104        &     104        &     104         &     104       \\
Individual FE           & Yes &   Yes  &  Yes &        Yes                        \\
Overall $R^2$               &       0.638         &                     &                     &                     \\
Root MSE            &       0.895         &       1.284         &       1.184         &       1.187         \\
First Stage F-Stat.                 &  & 23.09  &      727.33     &      358.98         \\
Hansen J stat.                   &                     &                     &                     &       0.624         \\
\toprule


 \\
                   &\multicolumn{4}{c}{{\bfseries [C] Split sample regression (June to August)}}\\\\

&\multicolumn{1}{c}{(1)}&\multicolumn{1}{c}{(2)}&\multicolumn{1}{c}{(3)}&\multicolumn{1}{c}{(4)}\\
                    &\multicolumn{1}{c}{FE}&\multicolumn{1}{c}{IV}&\multicolumn{1}{c}{IV}&\multicolumn{1}{c}{IV}\\
\cline{2-5}
$\mbox{Mobility range}_{i(t-1)}$&      -0.732\sym{***}&      -0.192         &      -0.371\sym{*}  &      -0.360         \\
                    &     (0.220)         &     (0.626)         &     (0.222)         &     (0.219)         \\
[1em]
Constant            &       8.863\sym{***}&                     &                     &                     \\
                    &     (0.026)         &                     &                     &                     \\
\hline
Observations        &         312         &         312         &         312         &         312         \\
Number Ids                &     104        &     104       &     104        &     104     \\
Individual FE           & Yes &   Yes  &  Yes &        Yes                        \\
Overall $R^2$              &       0.025         &                     &                     &                     \\
Root MSE            &       0.388         &       0.482         &       0.478         &       0.479         \\
First Stage F-Stat.                 &  &  24.65 &      340.96     &          196.74     \\
Hansen J stat.                   &                     &                     &                     &       0.087         \\
\hline\hline

\multicolumn{5}{l}{\footnotesize Robust standard errors in parentheses}\\
\multicolumn{5}{l}{\footnotesize \sym{*} \(p<0.10\), \sym{**} \(p<0.05\), \sym{***} \(p<0.01\)}\\
\end{tabular}}
\label{CIGfullsplit}
\end{table}

\begin{table}[H]
\centering
\def\sym#1{\ifmmode^{#1}\else\(^{#1}\)\fi}
\caption{Wage Guarantee Fund IV panel results}
\resizebox{11cm}{!}{
\begin{tabular}{l*{3}{c}}

\\ 
                   &\multicolumn{3}{c}{{\bfseries [A] Full Sample First Stage IV}}\\\\ &\multicolumn{1}{c}{(2)}&\multicolumn{1}{c}{(3)}&\multicolumn{1}{c}{(4)}\\
                    
                    &\multicolumn{1}{c}{$\mbox{Mobility}$}&\multicolumn{1}{c}{$\mbox{Mobility}$}&\multicolumn{1}{c}{$\mbox{Mobility}$}\\
                     &\multicolumn{1}{c}{$\mbox{range}_{i(t-1)}$}&\multicolumn{1}{c}{$\mbox{range}_{i(t-1)}$}&\multicolumn{1}{c}{$\mbox{range}_{i(t-1)}$}\\
\cline{2-4}
$\mbox{Betweenness}_{i(t-1)}$&      -1.426\sym{***}&                     &      -0.329\sym{**} \\
                    &     (0.203)         &                     &     (0.159)         \\
[1em]
Lockdown            &      -0.249\sym{***}&      -0.068\sym{***}&      -0.069\sym{***}\\
                    &     (0.017)         &     (0.013)         &     (0.013)         \\
[1em]
$\mbox{Share essentials}_{i(t-1)}$&                     &       0.017\sym{***}&       0.017\sym{***}\\
                    &                     &     (0.001)         &     (0.001)         \\
\hline
Observations        &         618         &         619         &         618         \\
Number Ids                &     104       &     104 &     104       \\
Root MSE                &       0.203         &       0.130         &       0.130         \\
Individual FE                &                    Yes &        Yes             &          Yes           \\
\toprule


\\
&\multicolumn{3}{c}{{\bfseries [B] Full Sample Reduced form IV}}\\\\ &\multicolumn{1}{c}{(2)}&\multicolumn{1}{c}{(3)}&\multicolumn{1}{c}{(4)}\\
                
                    &\multicolumn{1}{c}{$\mbox{ln WGF FTE}_{it}$}&\multicolumn{1}{c}{$\mbox{ln WGF FTE}_{it}$}&\multicolumn{1}{c}{$\mbox{ln WGF FTE}_{it}$}\\
\cline{2-4}
$\mbox{Betweenness}_{i(t-1)}$&       4.641\sym{***}&                     &      -1.336         \\
                    &     (1.395)         &                     &     (1.412)         \\
[1em]
Lockdown            &      -0.239\sym{*}  &      -1.217\sym{***}&      -1.223\sym{***}\\
                    &     (0.126)         &     (0.149)         &     (0.149)         \\
[1em]
$\mbox{Share essentials}_{i(t-1)}$&                     &      -0.092\sym{***}&      -0.093\sym{***}\\
                    &                     &     (0.007)         &     (0.007)         \\
\hline
Observations        &         618         &         619         &         618         \\
Number Ids                &     104        &     104     &     104         \\
Root MSE                &       1.528         &       1.269         &       1.269         \\
Individual FE                &                    Yes &        Yes             &          Yes           \\
\hline\hline
\multicolumn{4}{l}{\footnotesize Robust standard errors in parentheses}\\
\multicolumn{4}{l}{\footnotesize \sym{*} \(p<0.10\), \sym{**} \(p<0.05\), \sym{***} \(p<0.01\)}\\
\end{tabular}}
\label{CIGfirstred}
\end{table}

\begin{figure}[H]
\centering
\caption{NUTS 3 region-specific fixed effects estimates}%
\includegraphics[width=.5\textwidth]{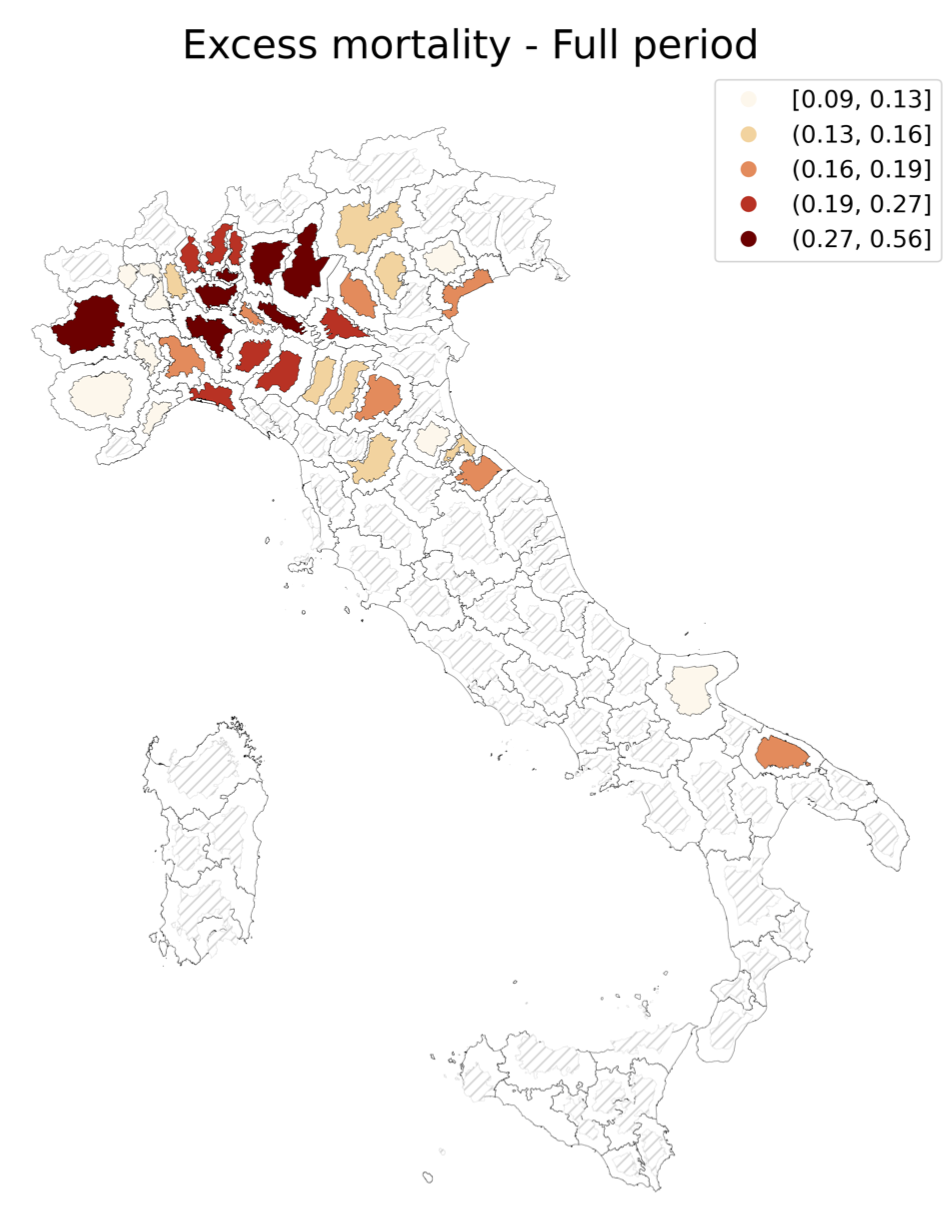}
\includegraphics[ width=.5\textwidth]{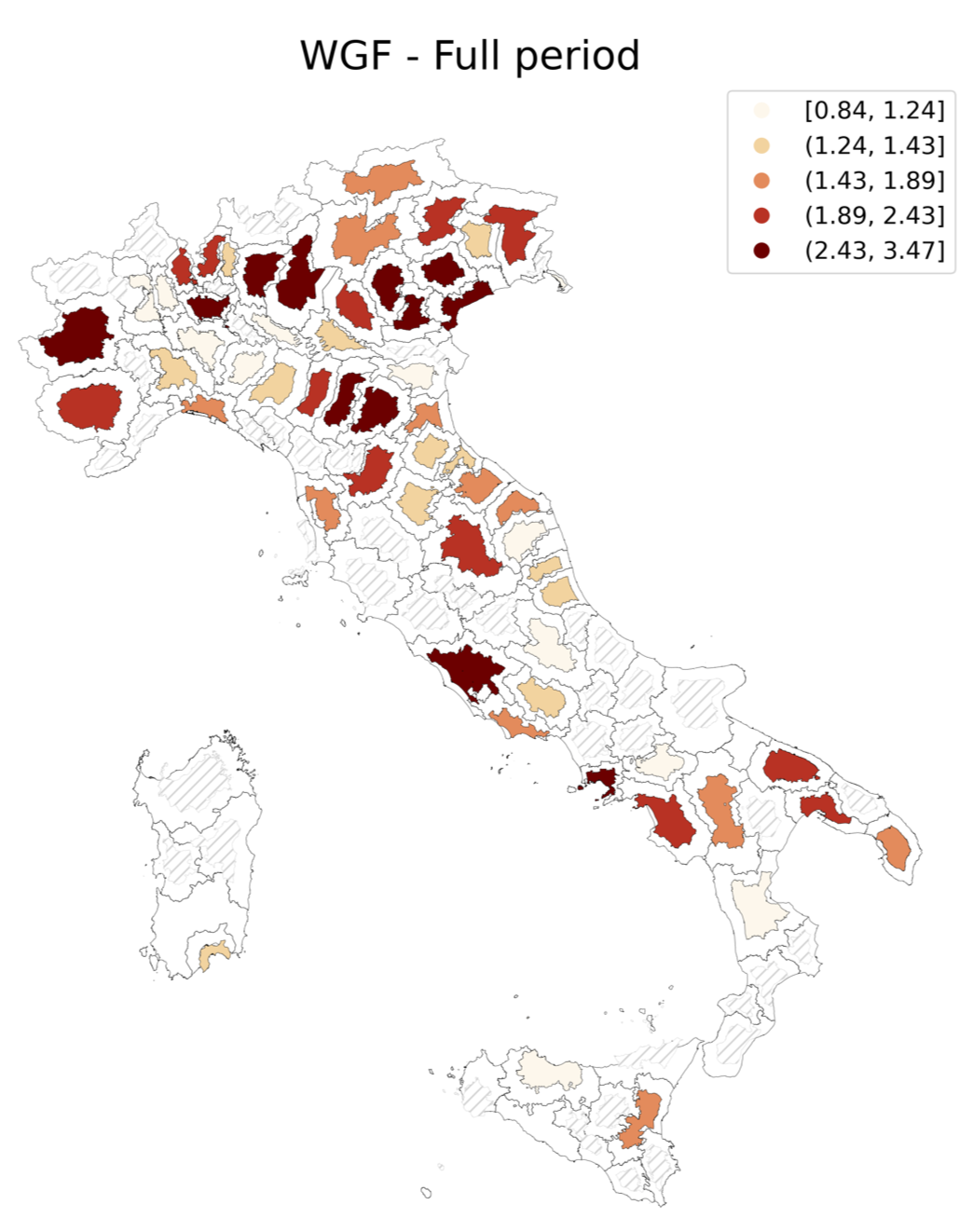}
\caption*{\footnotesize{Notes: the map shows a graphical representation of the NUTS 3 region-specific fixed effects estimated through equation $ln(\mbox{WGF FTE})_{it}  =  \beta_0 + \beta_1 Mob.Range_{i(t-1)} + \delta Lockdown_{t} + \sum_{j=2}^{N} pv_j d_{j,it} + \varepsilon_{it}$, when instrumenting Mobility Range by both NUTS 3 regions' centrality and the share of essential residents (IV model, specification (4), section [A], tables \ref{mortfullsplit} and \ref{CIGfullsplit}). The color intensity of NUTS 3 regions is proportional to coefficients $pv_j$. Each region in the plot is rescaled according to the average change in mobility occurred between March and April 2020.}}
\label{mapsFE}
\end{figure}

Our results have two main implications for the vaccine roll-out strategy. When moving to vaccinate the healthy share of the population, essential workers and workers not eligible for remote working should be prioritized since they increase mobility, thus inducing higher excess mortality. Second, based on the results of our analysis, we propose to prioritize the active share of the population based on unemployment risk.

This strategy implies to assign more vaccines to areas identified according to the estimates of the individual-specific fixed effects as from the full-sample IV model (specification [A](4), table \ref{CIGfullsplit}), considering the Wage Guarantee Fund as the dependent variable. Fixed effects coefficients are graphically represented in figure \ref{mapsFE}, where the second panel shows which territorial units experienced a higher increase in the Wage Guarantee Fund (FTE units).
Fixed effects estimates account for time-invariant NUTS 3 regions' characteristics, like demographic and socio-economic ones (which reasonably remain stable in the period we consider). 

The NUTS 3 regions that have been most in need for wages supplementation schemes have been identified according to fixed effects estimates performed on the entire period, spanning from March to August 2020 and on the lockdown time only (March to May). We assume that stricter restrictions are likely to be enforced in the months when the last steps of the campaign are about to start. Instead, we could refer to the estimates obtained when focusing on the post-lockdown period (June to August) if we expect milder (or almost absent) restrictions to be enforced.

\begin{figure}[t]
    \centering
    \caption{Difference in ranking $\Delta_i^{WGF}$}%
    \includegraphics[width=.80\textwidth]{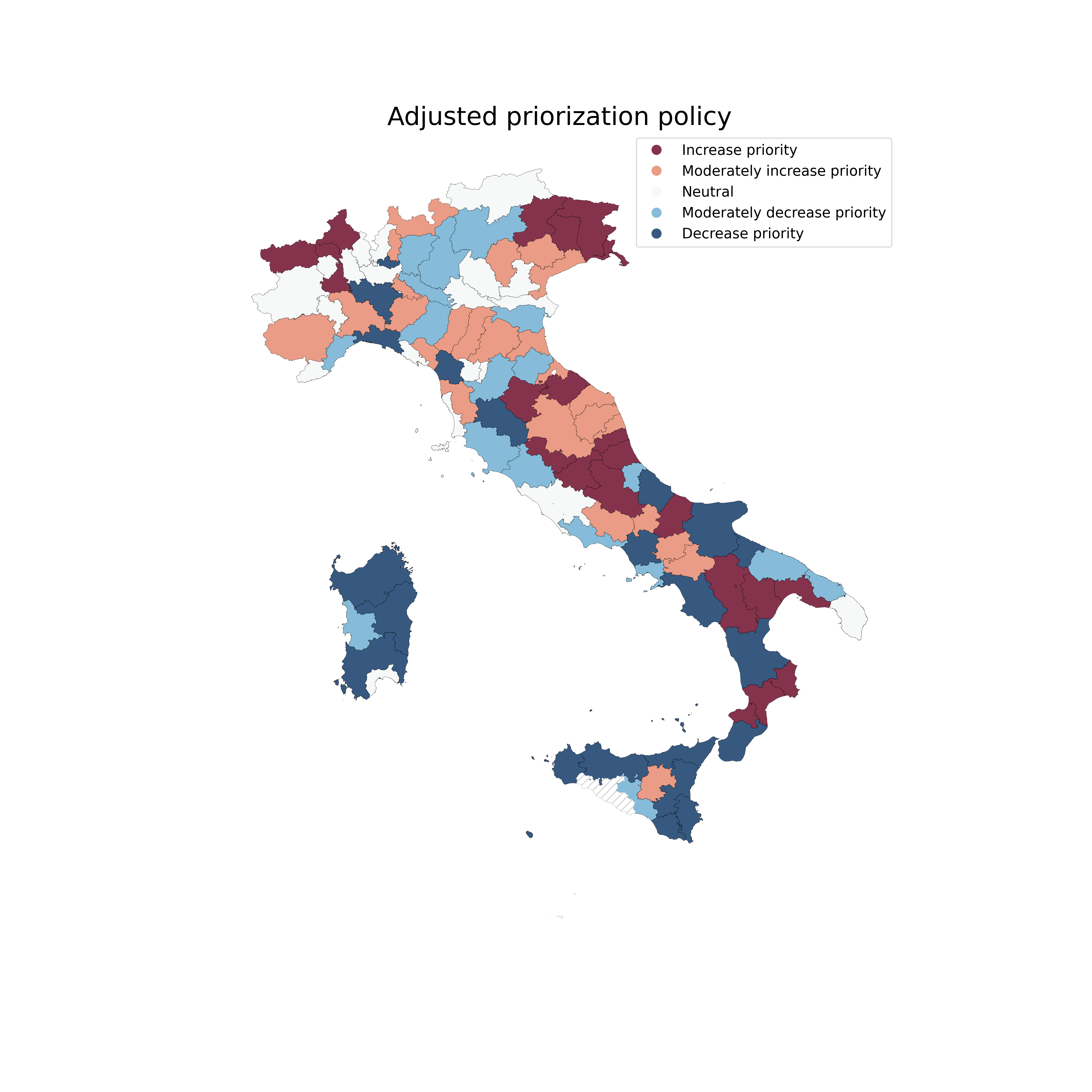}%
    \label{fig:diffmap}%
\end{figure}

We compare our allocation criterion with a benchmark based on the share of the working population\footnote{The number of people employed in region $i$ has been obtained by multiplying the 2019 value of the employment rate of people aged 20 to 64 as from ISTAT (Labour Force Survey), by the number of residents aged between 20 and 64.}.
 
Each NUTS 3 region is ranked according to the two criteria explained above, and we indicate with $R^{WP}_i$ and $R^{WGF}_i$  the position of region $i$ in the Working Population and Wage Guarantee Fund rankings respectively. To highlight possible inequalities, we compare the criteria by computing

\begin{equation}
    \Delta_i^{WGF}= R^{WGF}_i - R^{WP}_i
\end{equation}
\noindent where $\Delta_i{WGF}$ is the difference in the ranking positions between $\left\{WGF,EM\right\}$ and the working force ranking.
The distribution of $\Delta$  is reported in figure \ref{fig:diffmap}. The intensity of the color is proportional to $\Delta_i{WGF}$. Areas in light colors between blue and red tones are similar in both criteria ($\Delta \sim \pm 10$). NUTS 3 regions in red and dark red shades are those having $\Delta_i{WGF}<0$; hence they should be put in higher priority. Conversely, NUTS 3 regions in blue and dark blue shades ($\Delta>0)$, should be put in lower priority.

\section{Final discussion}

In this paper we analyze the impact of human mobility on excess mortality and furlough schemes in Italy. We assume that safe return-to-work will be possible for vaccinated workers, reactivating mobility and restoring full production capacity. This is because the negative health consequences of human mobility will be neutralized. Therefore, we propose a vaccine prioritization policy for the healthy and active share of the population in two stages. First, access to vaccination should be guaranteed to essential workers and workers not eligible for remote working. Then, return-to-work should be facilitated for the beneficiaries of wage guarantee schemes. This will be beneficial both in terms of a reallocation and more efficient use of public funds
and to reduce potential job losses. 
It is worth stressing that our recommendations refer to the last phase of the vaccination campaign when vulnerable categories according to the national strategic plan have already been vaccinated and immunized against the virus \citep{Sanita2020,Sanita2021b}. 

The proposed strategy puts in advantage those workers employed in the administrative areas in which wage integration measures have been used more intensively, allowing them to come back to a safe work\footnote{The prospected scenario does not take into account potential market labour flows (especially firings) which could occur when public policies issued to increase employment  protection, among which firing freeze, are lifted. The Law Decree "Sostegni", issued on 20 March 2021, 
has extended the firing freeze and the use of wage supplementation schemes for events related to the COVID-19 crisis. Employment terminations due to justified business reasons are suspended until the end of June 2021, and then, from July to the end of October, this measure will be enforced just as long as the employer (when eligible) is allowed to use derogatory wage guarantee schemes. 
}, triggering a gradual economic recovery. 
This policy's expected benefit can be interpreted chiefly in terms of a gradual resumption of most economic activities, freeing up public funds for long-term oriented social-protection programs\footnote{About the ongoing debate see \url{https://www.economist.com/leaders/2021/03/06/how-to-make-a-social-safety-net-for-the-post-covid-world?giftId=898c75eb-6951-47ce-93b1-499ff5852176}.}. 
We recall that, according to the European Commission \citep{EU2021}, the Italian government has committed around 19 billion euros to cover wage supplementation schemes\footnote{As from the same document, since March 2020  the Italian government has committed about 100.3 billion euros in accordance with three fiscal packages as from Law Decree no. 18 from 17 March, Law Decree no. 34 from 19 May, Law Decree no. 104 from 14 August, including, among the others, measures to support firms and employment.}, accounting for around $70\%$ of the total amount committed to employment support measures.\footnote{With the approval of the Law Decree "Sostegni", released on March 20$^{th}$ 2021,   
more funds have been committed to cover furlough schemes for events related to Covid-19: the use of the Ordinary Wage Guarantee Fund (Cassa integrazione guadagni ordinaria)  has been extended until the end of June 2021, while eligible employers can be allowed to derogatory wage guarantee schemes (Cassa integrazione guadagni in deroga)  until the end of December 2021. The suspension of employment agreements terminations for justified business reasons is in force as long as the firm is allowed to benefit from furlough schemes. 
}  

To support our proposal, we explored the link between the drop in mobility and the amount of the Wage Guarantee Fund expressed in full-time equivalent units, also providing  evidence on the association between changes in mobility 
and the number of excess deaths.


Results highlight a negative and significant relationship between mobility changes and the amount of the Wage Guarantee Fund (in full-time equivalent units) over March-August 2020. We find that a 1\% contraction in mobility (w.r.t. the baseline) explains a 5\% growth in the amount of the  Wage Guarantee Fund (FTE units) allowed in the following month. Looking at the interpretation of a full-time equivalent unit, a drop in human mobility explains an increase in the number of full-time working employees enrolled in wage guarantee schemes in the following month. 
The association becomes stronger when national lockdown was in force (March to May 2020), then gets milder and less significant after mobility restrictions have been loosened (June to August 2021).

Under the public health point of view, results show the existence of a positive and significant association between one month lagged mobility changes and the excess deaths recorded: a one percent drop in mobility (w.r.t. the baseline) explains a 0.5 percent drop in the number of excess deaths in the following month.

Our finding are in agreement with the literature, as a positive association between mobility changes and deaths has already been observed  by  \citet{Glaeser2020}, among others. In addition, \cite{Borri2020} highlighted a significant reduction in excess deaths (especially for older people) in those municipalities experiencing more restricting lockdown measures, then, the authors put in evidence how municipalities with a higher drop in the share of active people following business shutdowns are those showing a stronger contraction in mobility. Similarly, we notice that 
lower shares of essential working residents in a administrative region are associated to a higher mobility contraction. 
Moreover, more central regions are those experiencing a higher drop in mobility flows.\footnote{In line with this evidence, \citet{Krenz2020} detected a higher decline of mobility flows in less remote areas (lower travel time to the next urban center).} 


Our analysis points out an association between the share of people employed in essential industries and excess deaths going through human mobility flows. These results provide support for the inclusion of workers in essential activities among the priority categories.

Concerning the last stage of the vaccine delivery plan, we propose a prioritization criterion addressing the beneficiaries of furlough schemes 
and we test it against a benchmark based on the resident working force. 
Given the pandemic's heterogeneous impact on local employment \citep{Cerqua2020}, this criterion could be extended to include those workers who lost their job because of the pandemic crisis, to consider the whole active population with high long-term unemployment risk. 


Although we analyze the Italian case, our results are relevant for an international audience since European governments have issued similar employment protection measures as a response to the pandemic. Short-time working schemes meant to support the firms affected by the crisis have been introduced or extended in Europe \citep{EU2021}. European Union member states are allowed to ask for European funds to cover such employment protection measures: financial support in the form of loans granted on favorable terms is provided under the SURE instrument (temporary support to mitigate Unemployment Risks in an Emergency)\footnote{Italy is among those member states which will benefit  the most from the allocation of the resources provided under the SURE: around 27.4 billions out of the total amount of 90.3 billion euros approved by the European Council are gradually provided to Italy. Other member states which have been allowed to receive financial support under the SURE  are Belgium (7.8 billion), Spain (21.3 billion), Poland (11.2 billion), Portugal (5.9 billion), Greece (2.7 billion), Romania (4.1 billion). For further information see \url{https://ec.europa.eu/info/business-economy-euro/economic-and-fiscal-policy-coordination/financial-assistance-eu/funding-mechanisms-and-facilities/sure_en}}.  

As more data covering the period of the second epidemic wave and the effects of the COVID-19 crisis are to be released, future work will be devoted to a better characterization of the models developed and to a further refinement of the prioritization criteria for the vaccination campaign.


\section*{Acknowledgements}
AF is supported by SoBigData++ (EC grant n. 871042). AF VP and MR thank Francesca Santucci for help in comparing Facebook with Census data. AF VP and MR are grateful to Francesco Serti for valuable comments and suggestions. 

\appendix 

\section{Facebook and Italian census data}\label{app:facebook}
In this section we test how Facebook data represent a reliable approximation of the population commuting for work and study, i.e. excluding those that do not habitually move within and between the NUTS 3 areas. Our idea is to test this reliability by looking at the data provided by the Italian statistical office in 2011, i.e the last census data collection in Italy.

Both data are presented in the form of a origin destination matrix, where the links express the number of commuters who
travel between and within NUTS 3 regions. In the case of ISTAT data, people move for study or work, in all time slots, and data are averaged over a period of one year. On the other hand, Facebook data account for the people moving daily, sampled at 8 hours intervals. In order to compare data, we averaged over the whole available period the Facebook data obtaining an averaged origin destination matrix. Our hypothesis is that the temporal averaging leads to a origin destination matrix accounting for the habitual commuting of the study and workforce.

\begin{figure}[H]
\centering
\caption{Scatter plot of the commuting flows of ISTAT and Facebook mobility data (on the left). Focus on daytime (8AM-4PM): Scatter plot of the commuting flows of ISTAT and Facebook mobility data (on the right); fit 7.49, $R^2=0.97$.}
\includegraphics[scale=0.15]{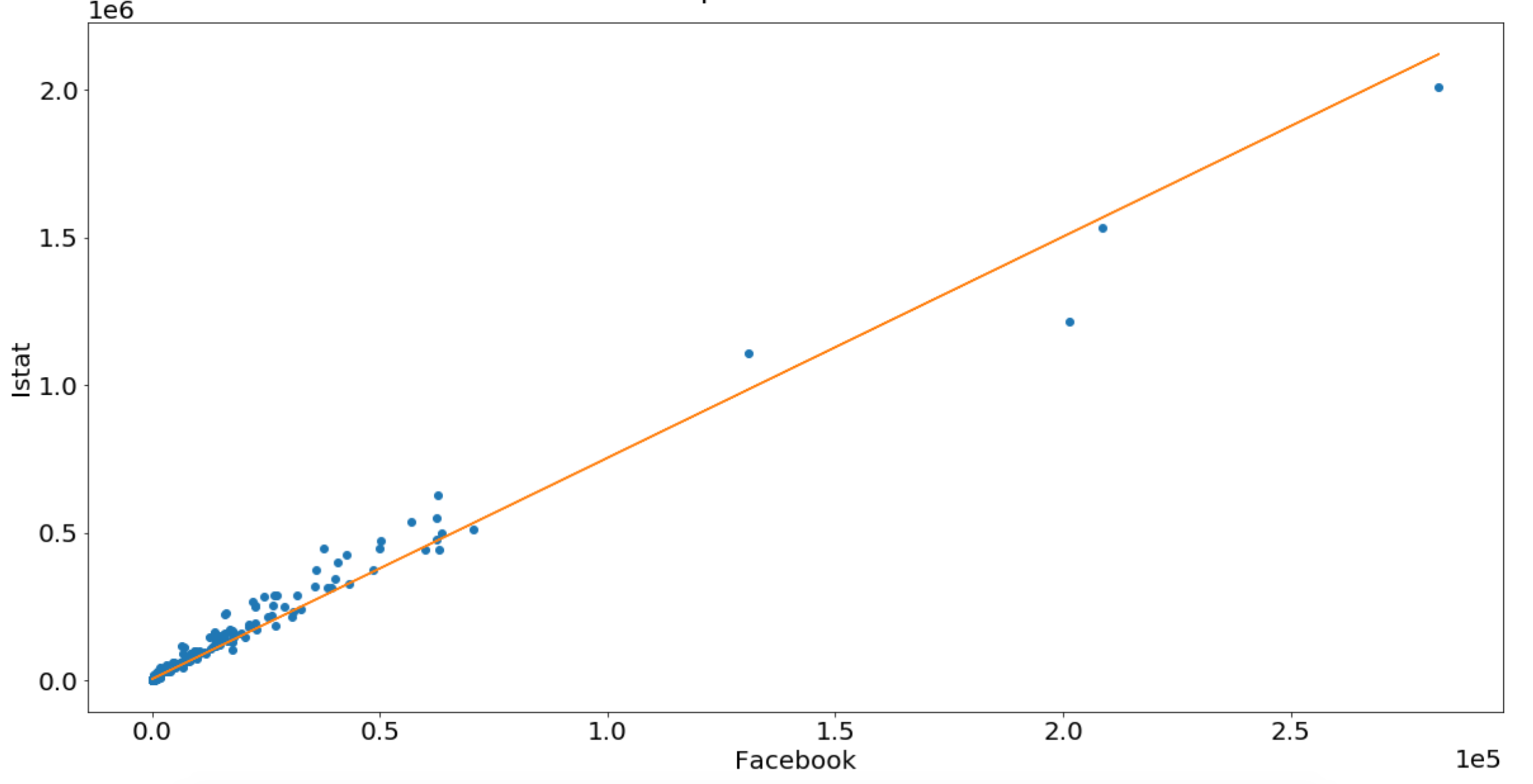}
\includegraphics[scale=0.16]{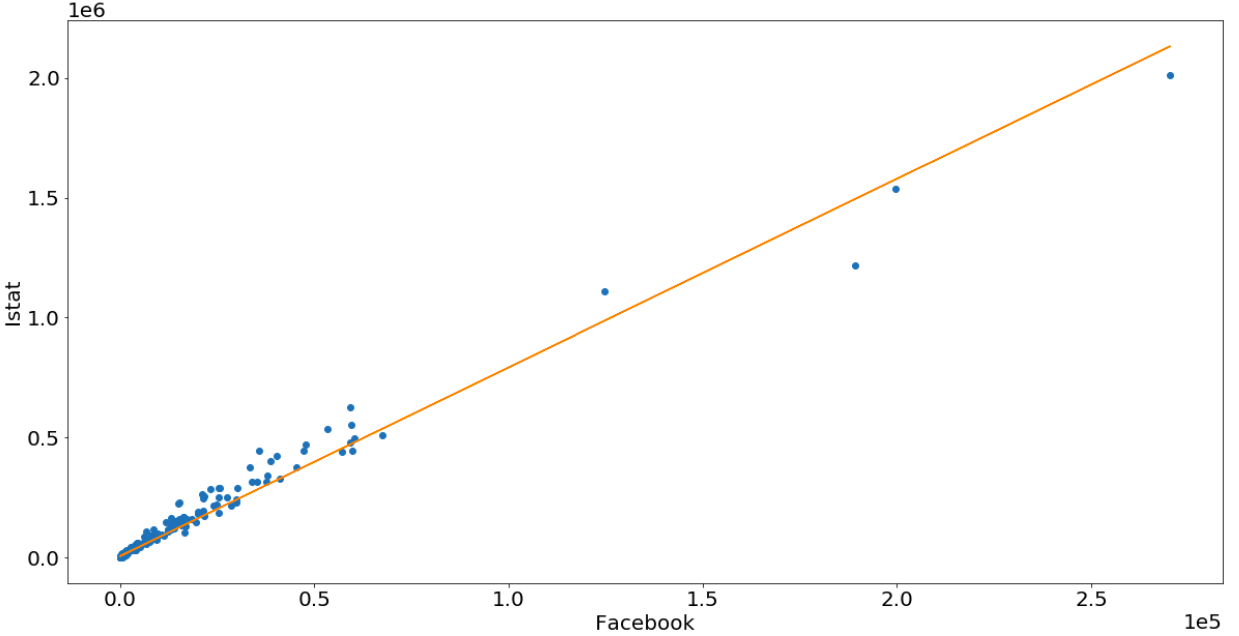}
\label{fig:ISTAT_scatter}
\end{figure}

Figure \ref{fig:ISTAT_scatter}(a) shows the plot of the weight of the links accounted in both
OD matrices. From a linear fit we find that Facebook data and ISTAT data are on average 1 to 7.5 ratio, confirming the data provided by \cite{Bonaccorsi2020}. As a further test, we considered only the Facebook traffic recorded during the daytime (8AM-4PM), comparing it to the corresponding period of ISTAT data. Results are depicted in Figure \ref{fig:ISTAT_scatter}(b), where the agreement of the two datasets is confirmed. According to these results we could consider, although sampling a smaller part of the population, Facebook movement data reliable under the point of view of the characterization of the mobility of habitual commuters.

\section{A robustness check: rainfall, mobility and furloughed workers}\label{app:robustness}
To corroborate our findings on the association between the Wage Guarantee Fund and mobility we performed a robustness check by employing an alternative instrumental variable.  


We estimated the IV model as in equation (\ref{IVmodel}) instrumenting for mobility by the share of rainy days recorded in a month\footnote{Daily weather data covering year 2020 have been collected from the weather forecast website \href{https://www.ilmeteo.it/portale/archivio-meteo/}{ilmeteo.it}.} over 2020 at Italian NUTS 3 level, namely {\itshape Rainfall share}. 

This model relies on the exogenous nature of weather shocks and on the assumption that mobility is the only channel through which rainfall could potentially have an impact on furlough schemes. 
Rainfall shocks have been used as an instrument in several scientific works investigating the relationship between economic shocks and conflicts \citep{Miguel2004,Sarsons2015,SandholtJensen2009,Bohlken2010}. 
A connection between weather shocks and income can be observed and theoretically justified for those countries whose economy depends largely on rain-fed agriculture as in the case of sub-Saharan African countries. However, the availability of developed irrigation infrastructures together with new agriculture technologies and a more relevant contribution of the industrial sector to the national (and regional) economy, make income growth less sensitive to weather shocks. 
In the specific case of Italy, we assume weather conditions to have a very negligible impact on economic activities, therefore we claim that the effect of rainfall on furlough schemes does not go through a potential shock on economic activities but through an induced variability in human mobility\footnote{As a further check the IV model has been estimated employing an alternative measure for the Wage Guarantee Fund, computed by excluding the share of allowed hours attributable to those sectors which could be more affected by a change in weather conditions, namely agriculture 
and food industry. The estimates are in line with those showed in table \ref{reg:cig_rainfall}, suggesting that the  effect of rainfall on furlough schemes does not go through a potential impact on economic activities. The output of this check is not displayed in the paper.}.  

Table \ref{reg:cig_rainfall} displays regression outputs: when instrumenting by {\itshape Rainfall share} we still observe that furlough schemes are negatively affected by mobility shocks, consolidating our previous findings.
Column (2) and (3) show split samples estimates obtained splitting the observation in "lockdown" and "post-lockdown" period: again, the association between mobility and furlough schemes appears to be stronger in the lockdown period before getting a bit weaker in the following months, nevertheless the estimates do not loose statistical significance. 

The full-sample first stage results (observations from March to August), column (1.a), show that mobility has been significantly affected by rainfall: an increase in the percentage of rainy days in a month explains a drop in mobility in the same period. 

\begin{table}[t]
\centering
\def\sym#1{\ifmmode^{#1}\else\(^{#1}\)\fi}
\caption{Wage Guarantee Fund and mobility: robustness}
\resizebox*{\textwidth}{!}{
\begin{tabular}{l*{5}{c}}

\\
 &\multicolumn{5}{c}{$\mbox{ln Wage Guarantee Fund FTE}_{it}$}\\\\
 \cline{2-6}
 \\
                    &\multicolumn{1}{c}{(1)}&\multicolumn{1}{c}{(1.a)}&\multicolumn{1}{c}{(1.b)}&\multicolumn{1}{c}{(2)}&\multicolumn{1}{c}{(3)}\\
                    &\multicolumn{1}{c}{Full Sample}&\multicolumn{1}{c}{Full Sample}&\multicolumn{1}{c}{Full Sample}&\multicolumn{1}{c}{March-May}&\multicolumn{1}{c}{June-August}\\
                    &\multicolumn{1}{c}{IV}&\multicolumn{1}{c}{First Stage}&\multicolumn{1}{c}{Reduced Form}&\multicolumn{1}{c}{IV}&\multicolumn{1}{c}{IV}\\
\cmidrule(lr){2-4}
\cmidrule(lr){5-6}
$\mbox{Mobility range}_{i(t-1)}$&     -11.136\sym{***}& & &     -12.884\sym{***}&      -7.706\sym{***}\\
                    &     (1.145)    & &      &     (0.981)         &     (2.777)         \\
[1em]
Lockdown            &      -3.089\sym{***}  &      -0.273\sym{***}&      -0.045         &                  &                     \\
                    &     (0.335)         &  (0.016) & (0.111) &                     &                     \\
[1em]
$\mbox{Rainfall share}_{i(t-1)}$&    &  -0.426\sym{***}&       4.749\sym{***} & & \\
                    &  &   (0.065)         &     (0.485)  & &       \\
\hline
Observations        &         611   &         611         &         611            &         302         &         308         \\
Number Ids                &     104   &     104         &     104      &     103         &     103         \\
Root MSE            &       1.502       &       0.202         &       1.423         &       2.018         &       1.183         \\
First Stage F-Stat.            &     42.58         &                 &                &         69.20       &         8.16       \\
Individual FE            &       Yes       &       Yes         &       Yes         &       Yes         &       Yes         \\
\hline\hline
\multicolumn{6}{l}{\footnotesize Robust standard errors in parentheses.}\\
\multicolumn{6}{l}{\footnotesize Columns (1.a) and (1.b) display estimates from the first stage and reduced from }\\
\multicolumn{6}{l}{\footnotesize
regressions on the full sample, which comprises observations from March to August.}\\
\multicolumn{6}{l}{\footnotesize \sym{*} \(p<0.10\), \sym{**} \(p<0.05\), \sym{***} \(p<0.01\)}\\
\end{tabular}}
\label{reg:cig_rainfall}
\end{table}

\section{Maps and plots}\label{app:maps}

\begin{figure}[H]
    \centering
    \caption{NUTS 3 region-specific fixed effects estimates: lockdown and post-lockdown periods}%
    \subfloat[\centering Excess Deaths: lockdown]{{\includegraphics[width=.4\textwidth]{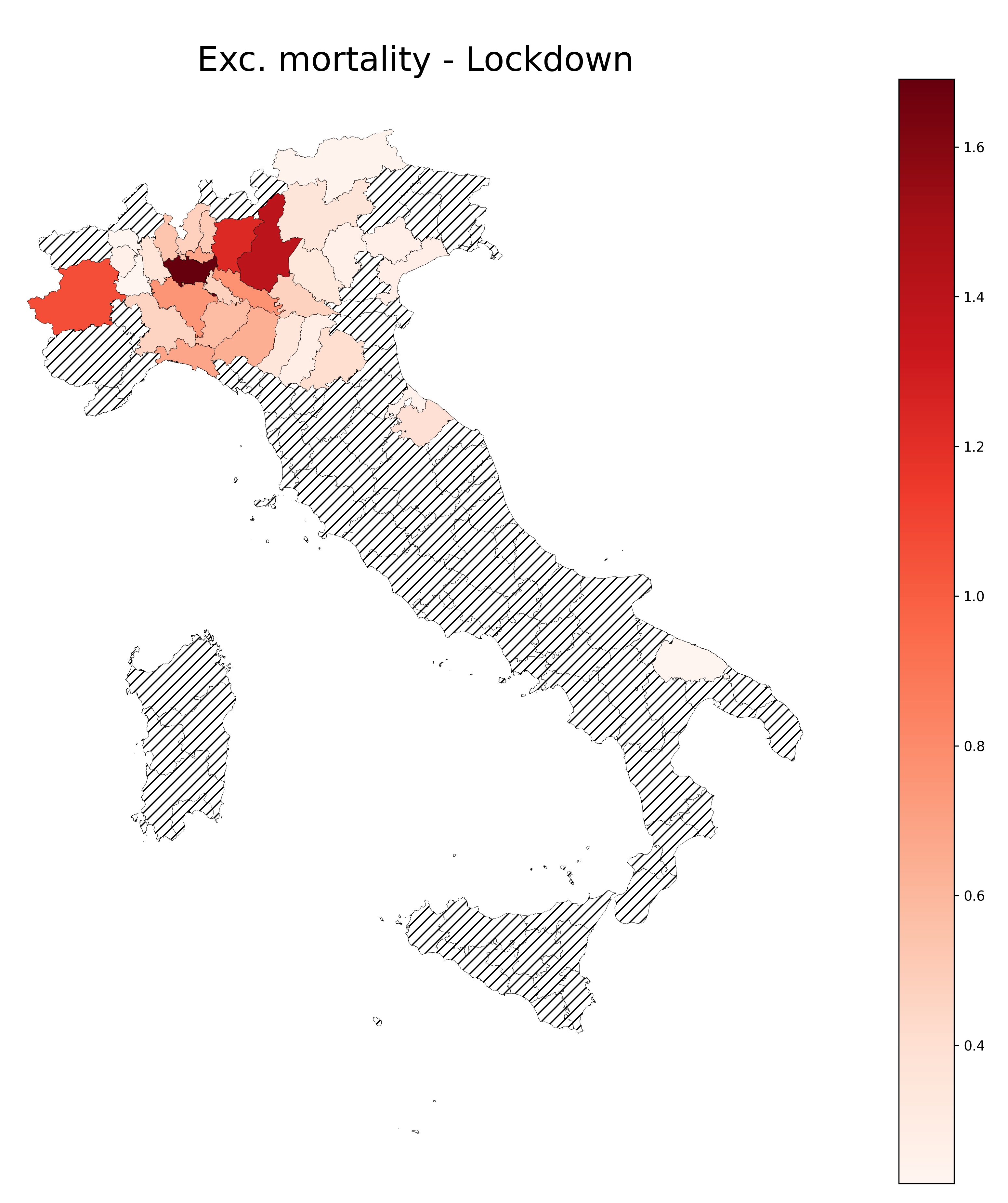} }}%
    \subfloat[\centering Wage Guarantee Fund: lockdown]{{\includegraphics[width=.4\textwidth]{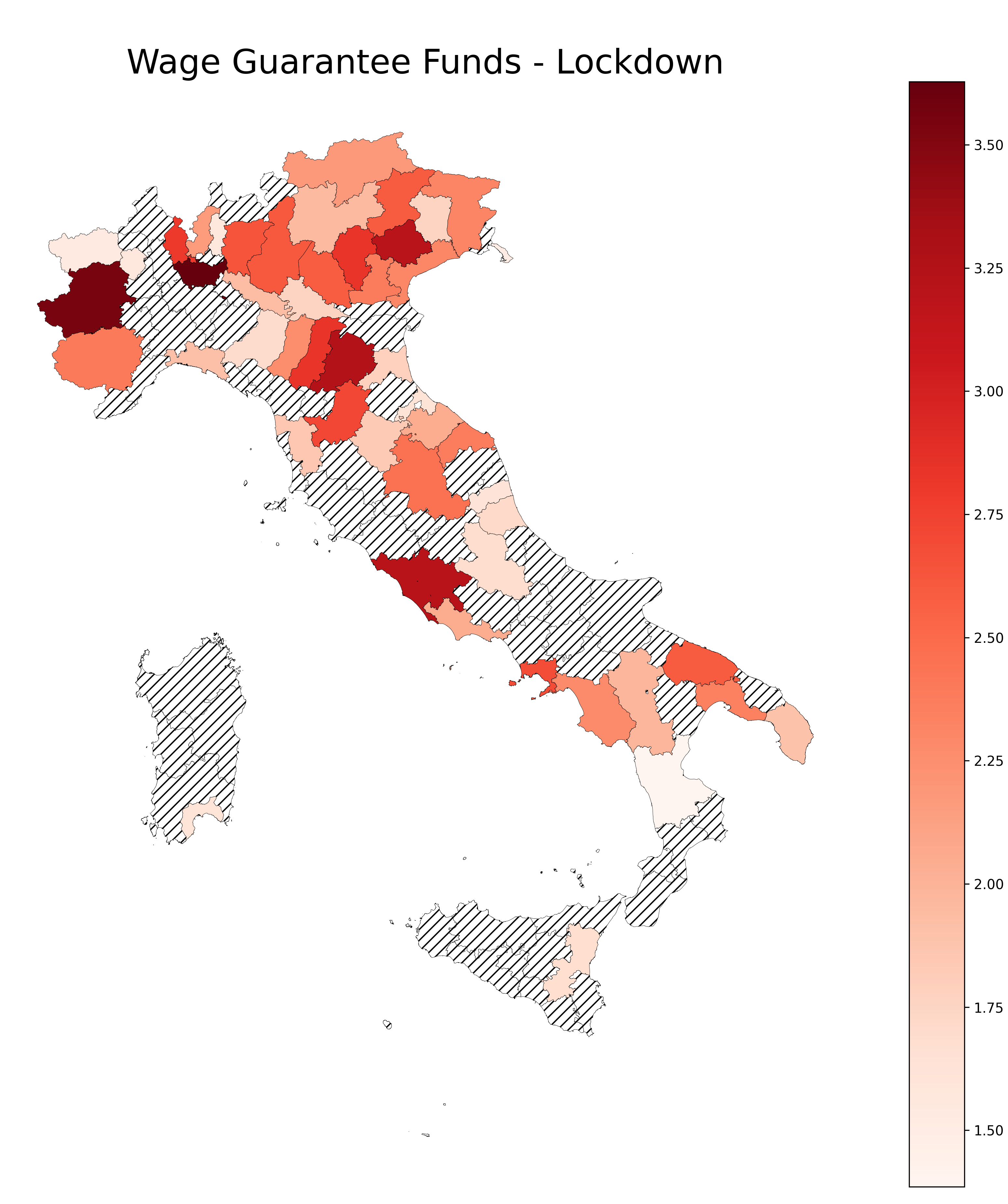} }}%
    \\
     \subfloat[\centering Excess Deaths: post-lockdown]{{\includegraphics[width=.4\textwidth]{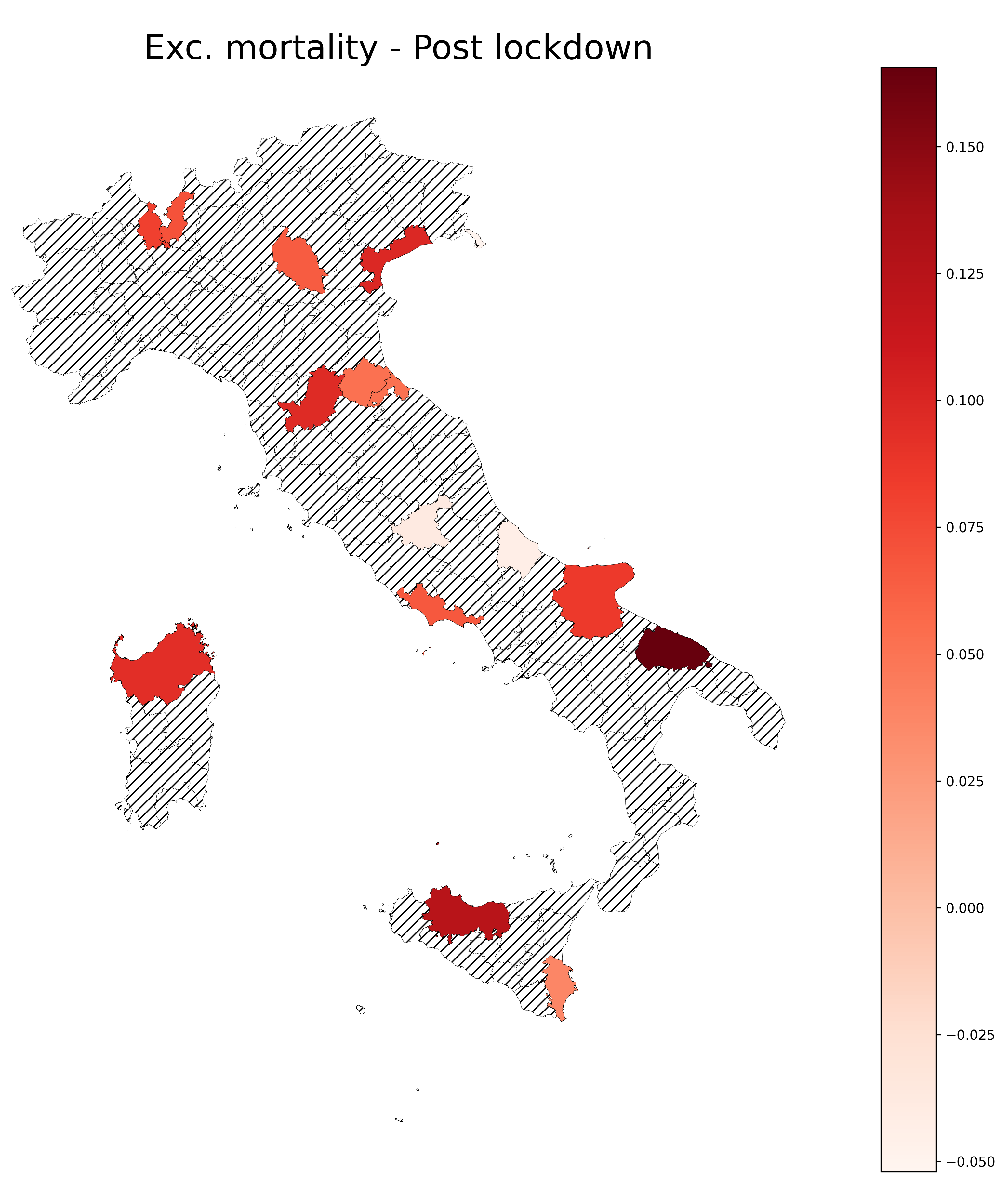} }}%
    \subfloat[\centering Wage Guarantee Fund: post-lockdown]{{\includegraphics[width=.4\textwidth]{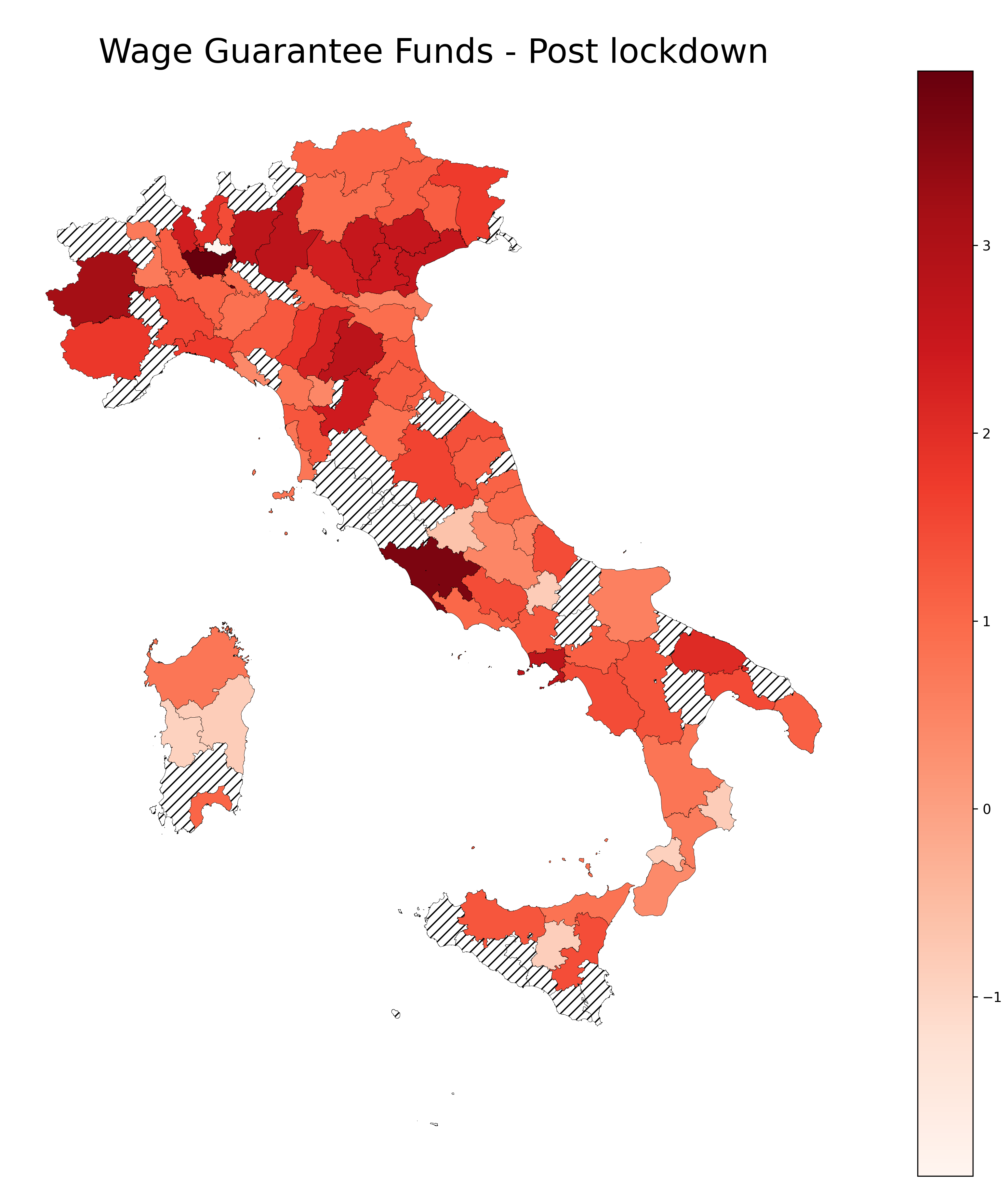} }}%
    \caption*{\footnotesize{Notes: the map shows a graphical representation of the individual fixed effects estimated through equation $ln(y)_{it}  =  \beta_0 + \beta_1 Mob.Range_{i(t-1)} + \sum_{j=2}^{N} pv_j d_{j,it} + \varepsilon_{it}$. Estimates from the lockdown and post-lockdown periods are represented.}}
    \label{fig:felockpost}%
\end{figure}

\begin{figure}[H]
    \centering
    \caption{Monthly Wage Guarantee Fund allowed hours}%
    \subfloat[\centering January]{{\includegraphics[width=.45\textwidth]{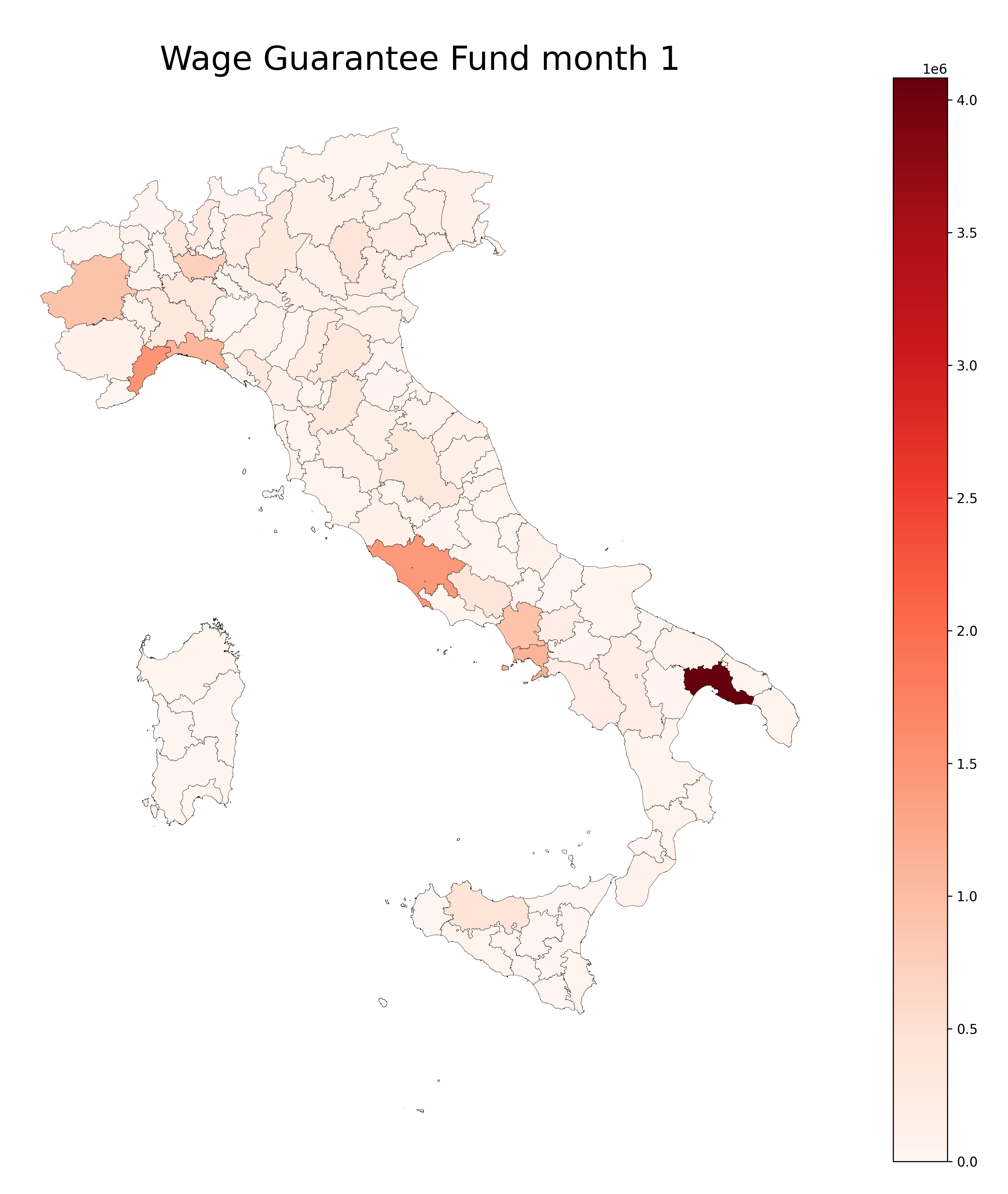} }}%
    \subfloat[\centering February]{{\includegraphics[width=.45\textwidth]{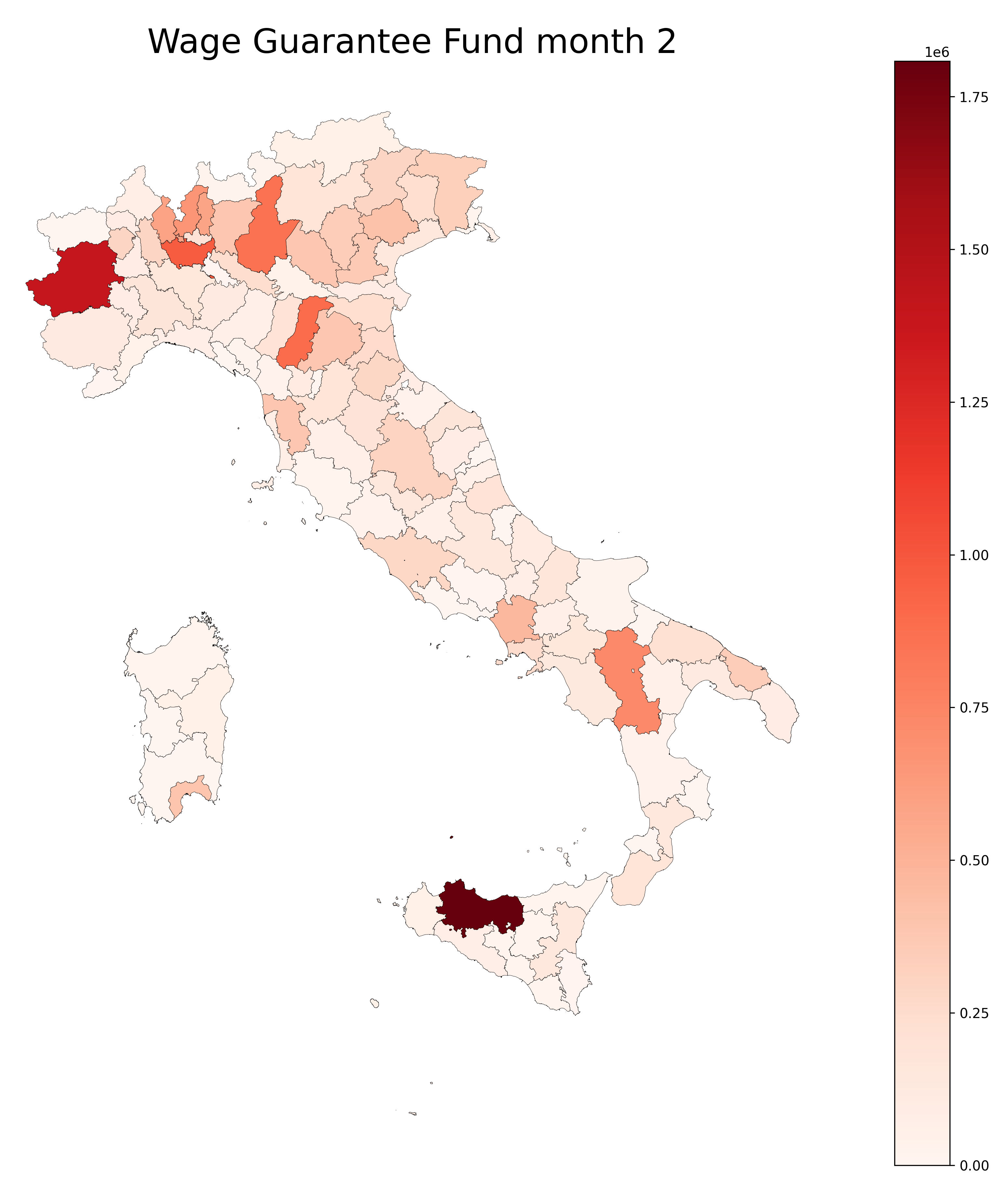} }}%
    \\
    \subfloat[\centering March]{{\includegraphics[width=.45\textwidth]{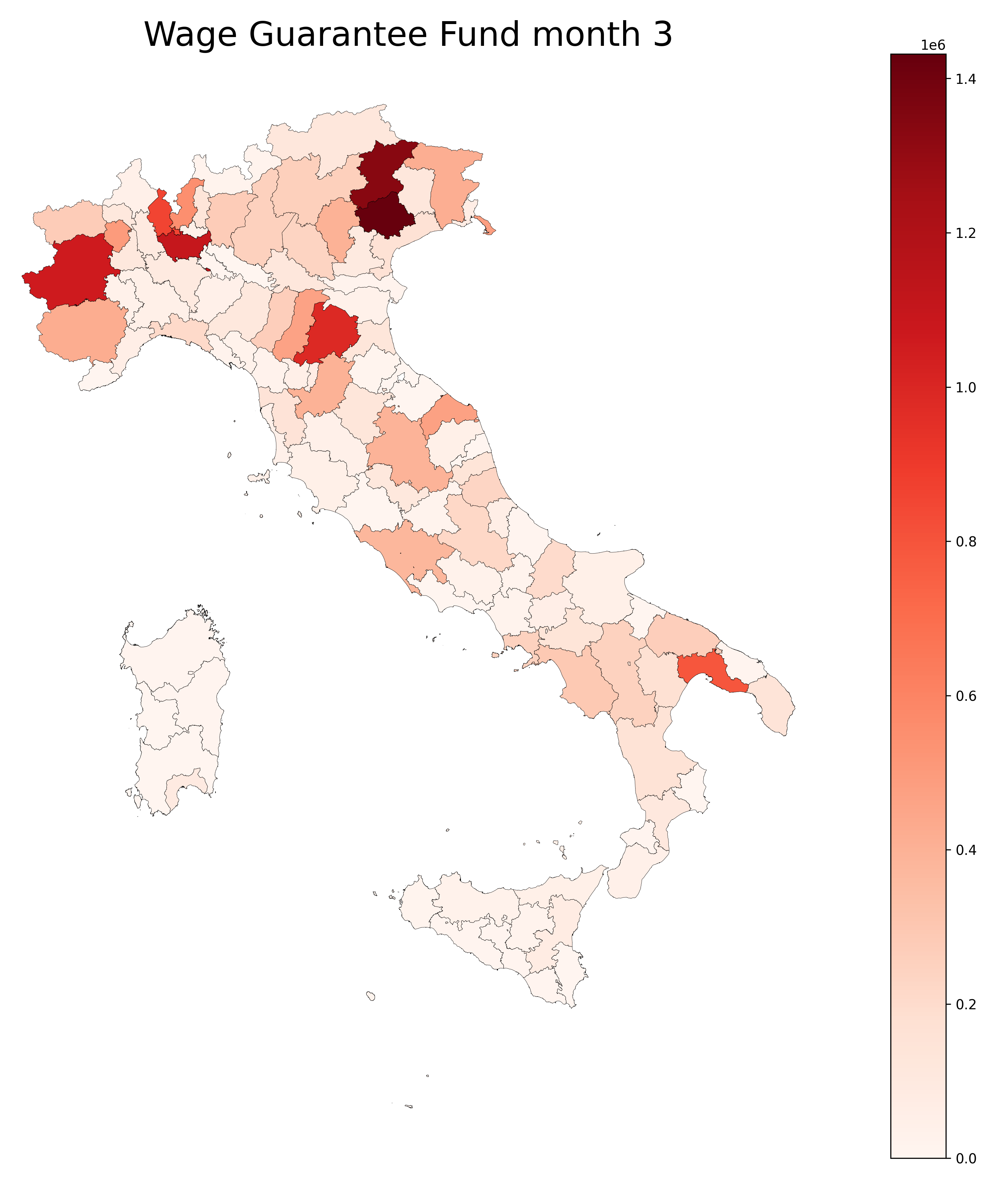} }}%
    \subfloat[\centering April]{{\includegraphics[width=.45\textwidth]{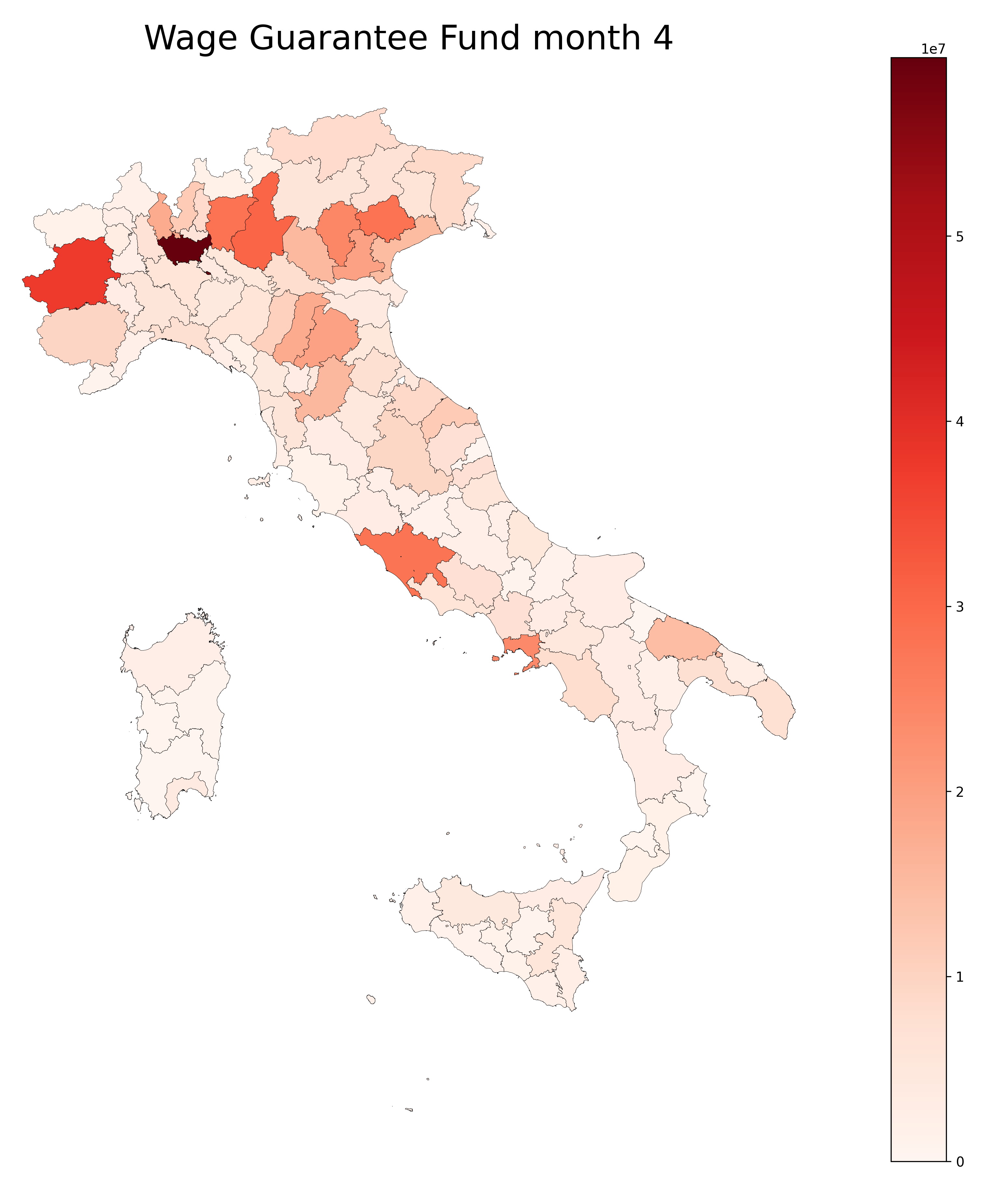} }}%

    \caption*{\footnotesize{}}
    \label{fig:mappeorecig}%
\end{figure}

\begin{figure}[H]
    \centering
    \subfloat[\centering May]{{\includegraphics[width=.45\textwidth]{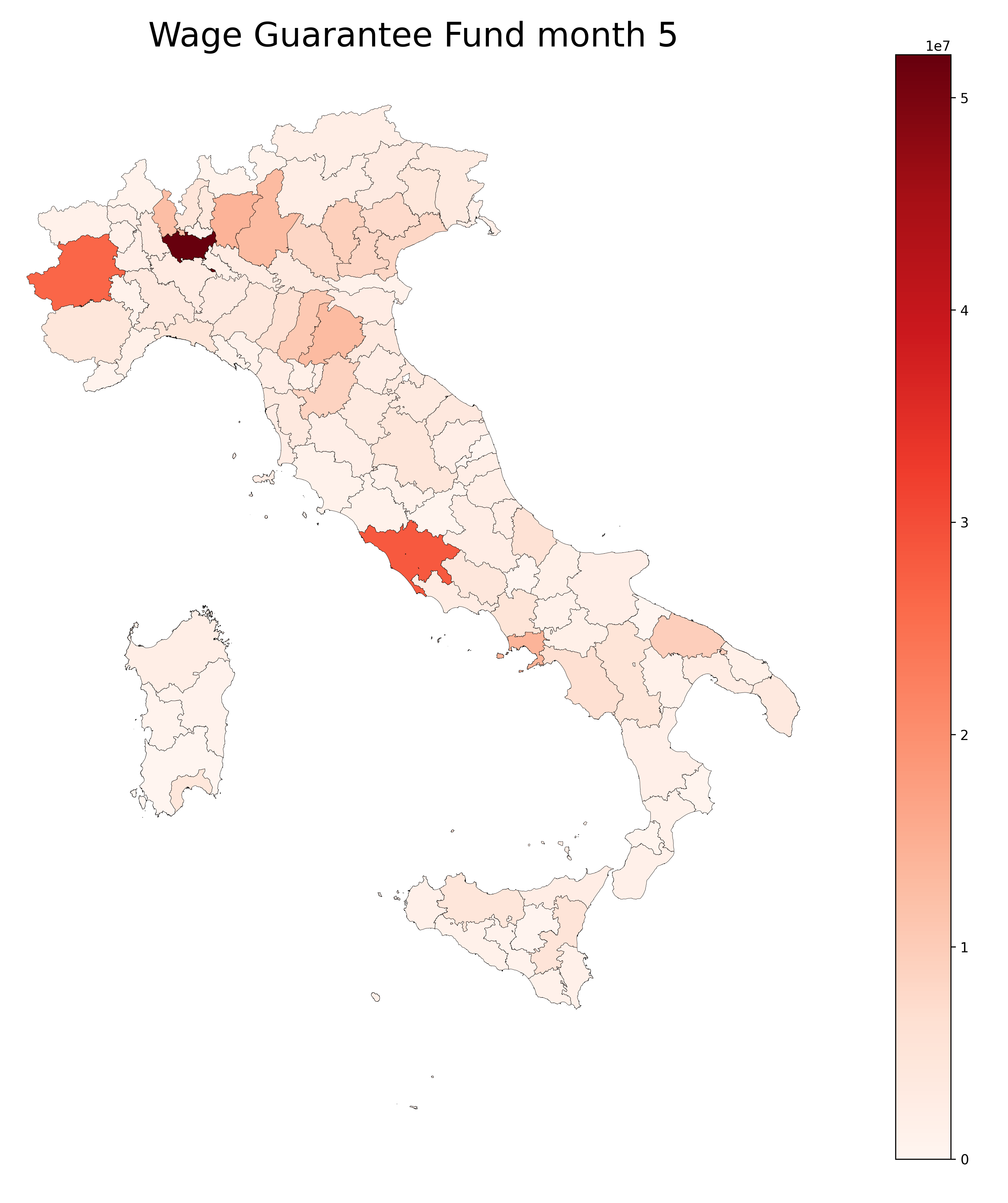} }}%
    \subfloat[\centering June]{{\includegraphics[width=.45\textwidth]{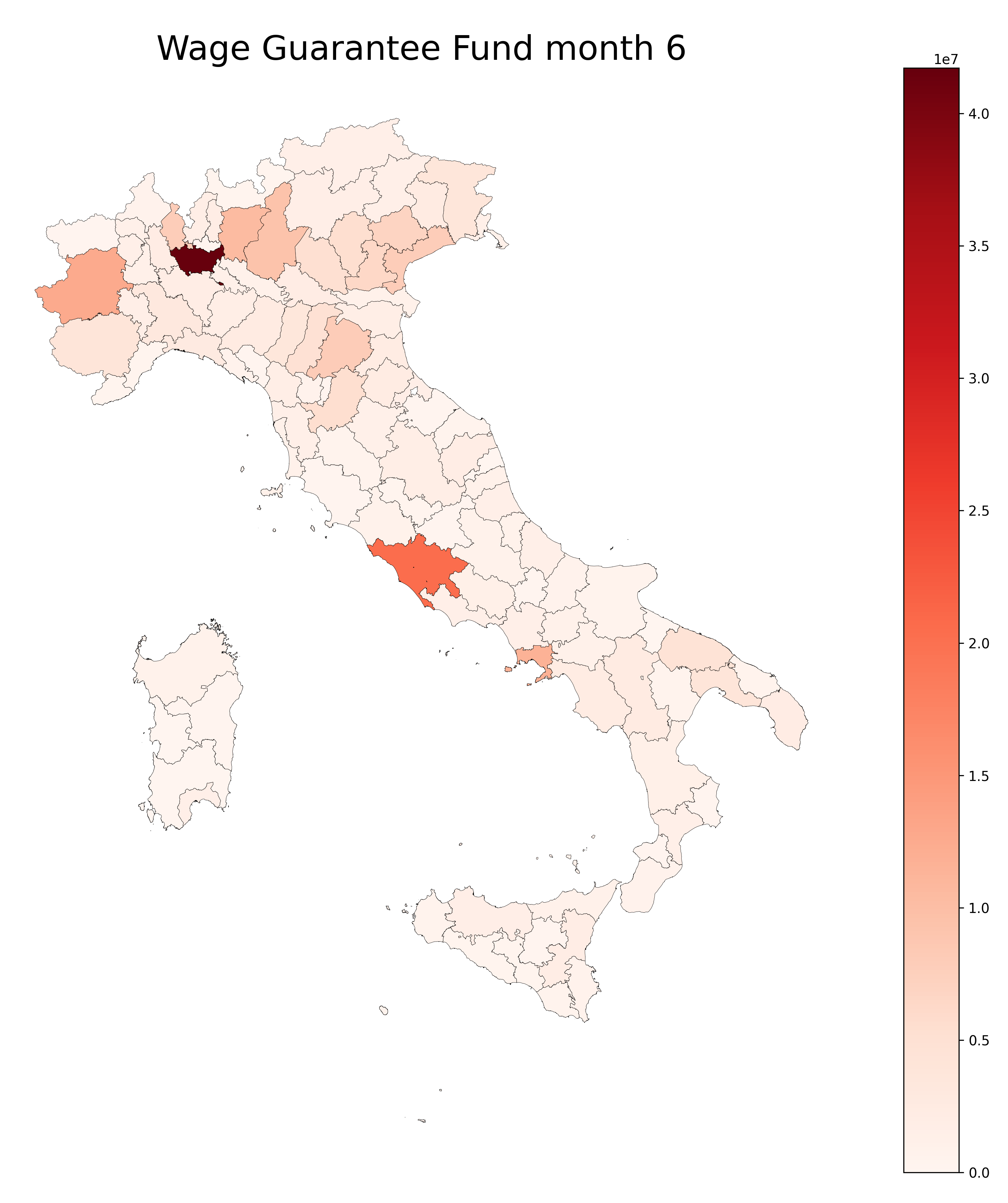} }}%
    \\
    \subfloat[\centering July]{{\includegraphics[width=.45\textwidth]{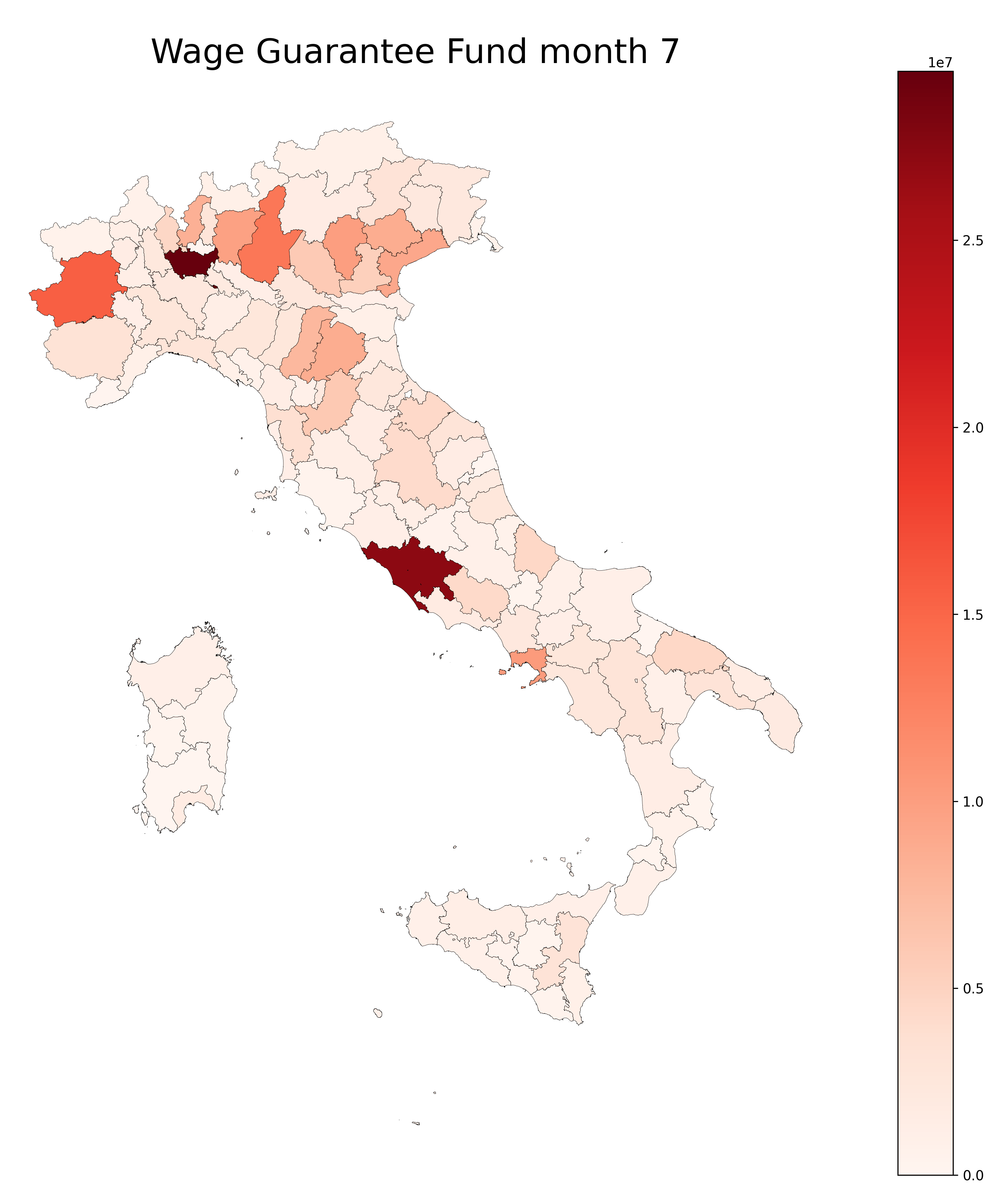} }}%
    \subfloat[\centering August]{{\includegraphics[width=.45\textwidth]{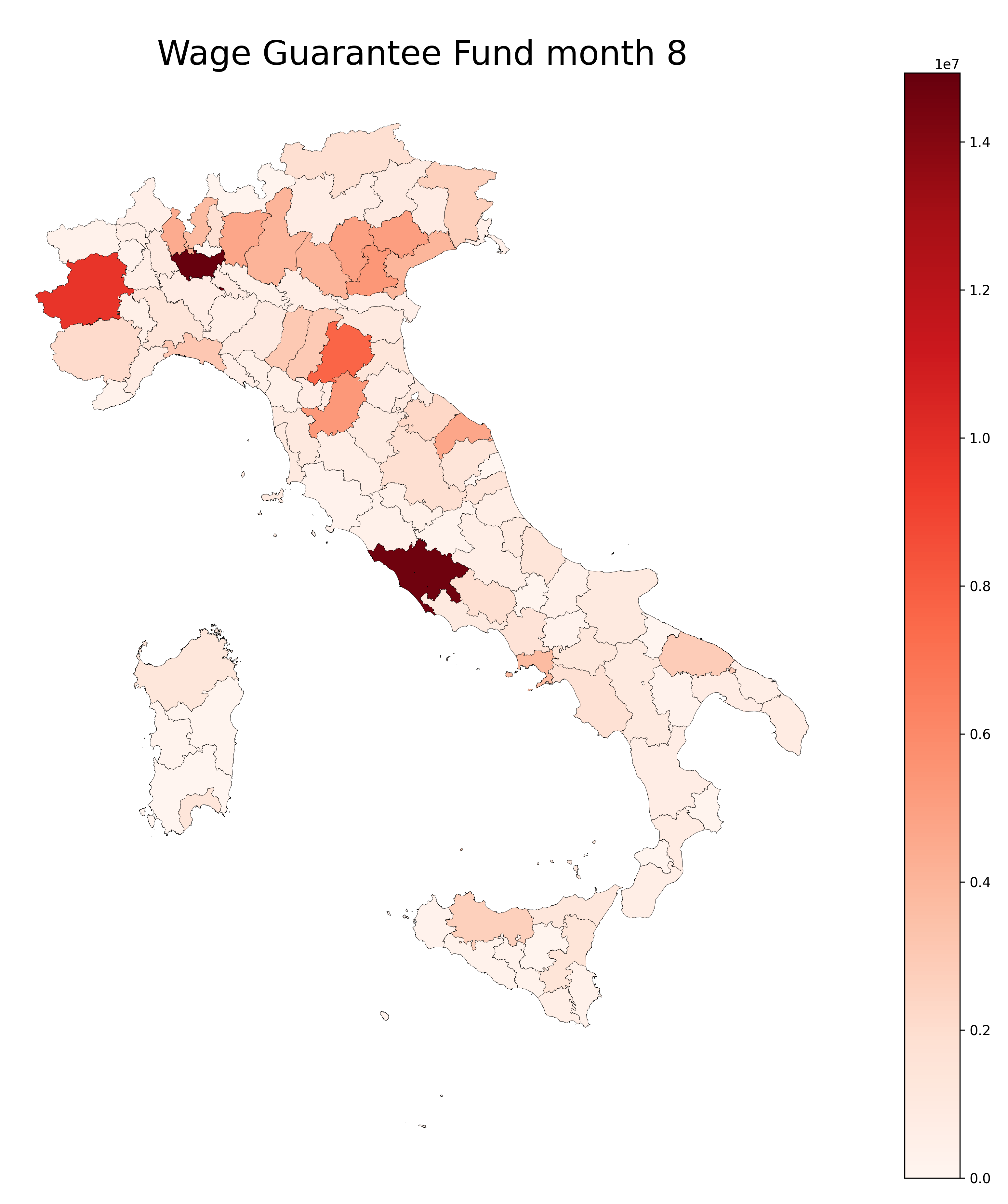} }}%
    \label{fig:mappeorecig2}%
\end{figure}


\newpage

\bibliographystyle{plainnat}

\bibliography{biblio}

\begin{thebibliography}{36}
\providecommand{\natexlab}[1]{#1}
\providecommand{\url}[1]{\texttt{#1}}
\expandafter\ifx\csname urlstyle\endcsname\relax
  \providecommand{\doi}[1]{doi: #1}\else
  \providecommand{\doi}{doi: \begingroup \urlstyle{rm}\Url}\fi

\bibitem[Acemoglu et~al.(2020)Acemoglu, Chernozhukov, Werning, and
  Whinston]{acemoglu2020multi}
D.~Acemoglu, V.~Chernozhukov, I.~Werning, and M.~D. Whinston.
\newblock A multi-risk sir model with optimally targeted lockdown.
\newblock Technical report, National Bureau of Economic Research, 2020.
\newblock URL \url{https://www.nber.org/papers/w27102}.

\bibitem[Ascani et~al.(2020)Ascani, Faggian, and Montresor]{Ascani2020}
A.~Ascani, A.~Faggian, and S.~Montresor.
\newblock The geography of covid-19 and the structure of local economies: The
  case of italy.
\newblock \emph{Journal of Regional Science}, 2020.
\newblock \doi{https://doi.org/10.1111/jors.12510}.

\bibitem[Besley and Stern(2020)]{BeselyStern2020}
T.~Besley and N.~Stern.
\newblock The economics of lockdown.
\newblock \emph{Fiscal Studies}, 41\penalty0 (3):\penalty0 493--513, 2020.
\newblock \doi{https://doi.org/10.1111/1475-5890.12246}.
\newblock URL
  \url{https://onlinelibrary.wiley.com/doi/abs/10.1111/1475-5890.12246}.

\bibitem[Bohlken and Sergenti(2010)]{Bohlken2010}
A.~T. Bohlken and E.~J. Sergenti.
\newblock Economic growth and ethnic violence: An empirical investigation of
  hindu—muslim riots in india.
\newblock \emph{Journal of Peace Research}, 47\penalty0 (5):\penalty0 589--600,
  2010.
\newblock \doi{10.1177/0022343310373032}.
\newblock URL \url{https://doi.org/10.1177/0022343310373032}.

\bibitem[Bonaccorsi et~al.(2020)Bonaccorsi, Pierri, Cinelli, Flori, Galeazzi,
  Porcelli, Schmidt, Valensise, Scala, Quattrociocchi, and
  Pammolli]{Bonaccorsi2020}
G.~Bonaccorsi, F.~Pierri, M.~Cinelli, A.~Flori, A.~Galeazzi, F.~Porcelli, A.~L.
  Schmidt, C.~M. Valensise, A.~Scala, W.~Quattrociocchi, and F.~Pammolli.
\newblock {Economic and social consequences of human mobility restrictions
  under COVID-19}.
\newblock \emph{Proceedings of the National Academy of Sciences of the United
  States of America}, 117\penalty0 (27):\penalty0 15530--15535, 2020.
\newblock ISSN 10916490.
\newblock \doi{10.1073/pnas.2007658117}.

\bibitem[Borri et~al.(2020)Borri, Drago, Santantonio, and Sobbrio]{Borri2020}
N.~Borri, F.~Drago, C.~Santantonio, and F.~Sobbrio.
\newblock {The 'Great Lockdown': Inactive Workers and Mortality by COVID-19}.
\newblock 2020.
\newblock URL
  \url{https://cepr.org/active/publications/discussion_papers/dp.php?dpno=15317}.

\bibitem[Borsati et~al.(2020)Borsati, Nocera, and Percoco]{Borsati2020}
M.~Borsati, S.~Nocera, and M.~Percoco.
\newblock {Questioning the spatial association between the spread of Covid-19
  and transit usage in Italy}.
\newblock 2020.

\bibitem[Buonanno et~al.(2020)Buonanno, Galletta, and Puca]{Buonanno2020}
P.~Buonanno, S.~Galletta, and M.~Puca.
\newblock {Estimating the severity of COVID-19: Evidence from the Italian
  epicenter}.
\newblock \emph{PLoS ONE}, 15\penalty0 (10 October):\penalty0 1--13, 2020.
\newblock \doi{10.1371/journal.pone.0239569}.
\newblock URL \url{http://dx.doi.org/10.1371/journal.pone.0239569}.

\bibitem[Casarico and Lattanzio(2020)]{Casarico2020}
A.~Casarico and S.~Lattanzio.
\newblock {The heterogeneous effects of COVID-19 on labor market flows:
  Evidence from administrative data}.
\newblock 2020.

\bibitem[Cerqua and Letta(2020)]{Cerqua2020}
A.~Cerqua and M.~Letta.
\newblock {Local economies amidst the COVID-19 crisis in Italy: A tale of
  diverging trajectories}.
\newblock 2020.

\bibitem[Chetty et~al.(2020)Chetty, Friedman, Hendren, and Stepner]{Chetty2020}
R.~Chetty, J.N. Friedman, N.~Hendren, and M.~Stepner.
\newblock {How did COVID-19 and stabilization policies affect spending and
  employment? A new real-time economic tracker based on private sector data}.
\newblock 2020.
\newblock URL \url{http://www.nber.org/papers/w27431\%0ANATIONAL}.

\bibitem[Commission(2021)]{EU2021}
European Commission.
\newblock {Policy measures taken against the spread and impact of the
  coronavirus - January 2021}.
\newblock Technical report, 2021.
\newblock URL
  \url{https://ec.europa.eu/info/sites/info/files/coronovirus_policy_measures_14_january_2021.pdf}.

\bibitem[della Sanit{\`{a}}(2020)]{Sanita2020}
Ministero della Sanit{\`{a}}.
\newblock {Piano strategico: Elementi di preparazione e di implementazione
  della strategia vaccinale}.
\newblock Technical report, 2020.
\newblock URL
  \url{http://www.salute.gov.it/portale/documentazione/p6_2_2_1.jsp?id=2986}.

\bibitem[della Sanit{\`{a}}(2021)]{Sanita2021b}
Ministero della Sanit{\`{a}}.
\newblock {Raccomandazioni ad interim sui gruppi target della vaccinazione
  anti-SARS-CoV-2/COVID-19}.
\newblock Technical report, 2021.
\newblock URL
  \url{http://www.salute.gov.it/imgs/C_17_pubblicazioni_3014_allegato.pdf}.

\bibitem[Fana et~al.(2020)Fana, P{\'e}rez, and
  Fern{\'a}ndez-Mac{\'\i}as]{fana2020employment}
Marta Fana, Sergio~Torrej{\'o}n P{\'e}rez, and Enrique
  Fern{\'a}ndez-Mac{\'\i}as.
\newblock Employment impact of covid-19 crisis: from short term effects to long
  terms prospects.
\newblock \emph{Journal of Industrial and Business Economics}, 47\penalty0
  (3):\penalty0 391--410, 2020.

\bibitem[Farboodi et~al.(2020)Farboodi, Jarosch, and Shimer]{Farboodi2020}
M.~Farboodi, G.~Jarosch, and R.~J. Shimer.
\newblock {Internal and External Effects of Social Distancing in a Pandemic}.
\newblock 2020.
\newblock URL \url{https://www.nber.org/papers/w27059}.

\bibitem[Favero et~al.(2020)Favero, Ichino, and Rustichini]{Favero2020}
C.~A. Favero, A.~Ichino, and A.~Rustichini.
\newblock {Restarting the Economy While Saving Lives Under COVID-19}.
\newblock 2020.
\newblock URL \url{http://dx.doi.org/10.2139/ssrn.3580626}.

\bibitem[Glaeser et~al.(2020)Glaeser, Gorback, and Redding]{Glaeser2020}
E.~L. Glaeser, C.~Gorback, and S.~J. Redding.
\newblock {JUE insight: How much does COVID-19 increase with mobility? Evidence
  from New York and four other U.S. cities}.
\newblock \emph{Journal of Urban Economics}, \penalty0 (July):\penalty0 103292,
  2020.
\newblock ISSN 00941190.
\newblock \doi{10.1016/j.jue.2020.103292}.

\bibitem[Goolsbee and Syverson(2021)]{Goolsbee2021}
A.~Goolsbee and C.~Syverson.
\newblock {Fear, lockdown, and diversion: Comparing drivers of pandemic
  economic decline 2020}.
\newblock \emph{Journal of Public Economics}, 193:\penalty0 104311, 2021.
\newblock ISSN 00472727.
\newblock \doi{10.1016/j.jpubeco.2020.104311}.

\bibitem[INPS(2020)]{INPS_CIG}
INPS.
\newblock Opendata portal, 2020.
\newblock URL \url{https://www.inps.it/OpenData/default.aspx?iidlink=103}.

\bibitem[INPS and d'Italia(2020)]{INPS2020}
INPS and Banca d'Italia.
\newblock {Le imprese e i lavoratori in cassa integrazione COVID nei mesi di
  marzo e aprile}, 2020.
\newblock URL
  \url{https://www.bancaditalia.it/pubblicazioni/note-covid-19/2020/Prime-evidenze-CIG_29072020.pdf}.

\bibitem[ISTAT(2020)]{ISTAT_mort}
ISTAT.
\newblock Decessi e cause di morte, 2020.
\newblock URL \url{https://www.istat.it/it/archivio/240401}.

\bibitem[Krenz and Strulik(2020)]{Krenz2020}
A.~Krenz and H.~Strulik.
\newblock {The Benefits of Remoteness - Digital Mobility Data, Regional Road
  Infrastructure, and COVID-19 Infection}.
\newblock 2020.
\newblock URL
  \url{https://www.econstor.eu/bitstream/10419/219321/1/1700585576.pdf}.

\bibitem[Maas et~al.(2019)Maas, Gros, McGorman, {Alex Dow}, Iyer, Park, and
  Nayak]{Maas2019}
P.~Maas, A.~Gros, L.~McGorman, P.~{Alex Dow}, S.~Iyer, W.~Park, and C.~Nayak.
\newblock {Facebook disaster maps: Aggregate insights for crisis response {\&}
  recovery}.
\newblock In \emph{Proceedings of the International ISCRAM Conference}, volume
  2019-May, pages 836--847, 2019.

\bibitem[Medlock and Galvani(2009)]{Medlock2018}
J.~Medlock and A.~P. Galvani.
\newblock Optimizing influenza vaccine distribution.
\newblock \emph{Science}, 325:\penalty0 1705, 2009.

\bibitem[Miguel et~al.(2004)Miguel, Satyanath, and Sergenti]{Miguel2004}
E.~Miguel, S.~Satyanath, and E.~Sergenti.
\newblock Economic shocks and civil conflict: An instrumental variables
  approach.
\newblock \emph{Journal of Political Economy}, 112\penalty0 (4):\penalty0
  725--753, 2004.
\newblock \doi{10.1086/421174}.
\newblock URL \url{https://doi.org/10.1086/421174}.

\bibitem[Muscillo et~al.(2021)Muscillo, Pin, and Razzolini]{muscillo2021}
Alessio Muscillo, Paolo Pin, and Tiziano Razzolini.
\newblock {Spreading of an infectious disease between different locations}.
\newblock \emph{Journal of Economic Behavior and Organization}, 183:\penalty0
  508--532, mar 2021.
\newblock ISSN 01672681.
\newblock \doi{10.1016/j.jebo.2021.01.004}.

\bibitem[Newman(2010)]{Newman}
M.E.J. Newman.
\newblock \emph{Networks: An introduction}.
\newblock Oxford University Press, 2010.
\newblock ISBN 978-0199206650.

\bibitem[Sah et~al.(2018)Sah, Medlock, Fitzpatrick, Singer, and
  Galvani]{Sah2018}
P.~Sah, J.~Medlock, M.~C. Fitzpatrick, B.~H. Singer, and A.P. Galvani.
\newblock Optimizing the impact of low-efficacy influenza vaccines.
\newblock \emph{Proc. Natl. Acad. Sci.}, 115:\penalty0 5151, 2018.

\bibitem[Sandholt~Jensen and Skrede~Gleditsch(2009)]{SandholtJensen2009}
P.~Sandholt~Jensen and K.~Skrede~Gleditsch.
\newblock Rain, growth, and civil war: The importance of location.
\newblock \emph{Defence and Peace Economics}, 20\penalty0 (5):\penalty0
  359--372, 2009.
\newblock \doi{10.1080/10242690902868277}.
\newblock URL \url{https://doi.org/10.1080/10242690902868277}.

\bibitem[Sarsons(2015)]{Sarsons2015}
H.~Sarsons.
\newblock Rainfall and conflict: A cautionary tale.
\newblock \emph{Journal of Development Economics}, 115:\penalty0 62--72, 2015.
\newblock \doi{https://doi.org/10.1016/j.jdeveco.2014.12.007}.
\newblock URL
  \url{https://www.sciencedirect.com/science/article/pii/S030438781400159X}.

\bibitem[Viviano(2020)]{Viviano2020}
E.~Viviano.
\newblock {Alcune stime preliminari degli effetti delle misure di sostegno sul
  mercato del lavoro}, 2020.
\newblock URL
  \url{https://www.bancaditalia.it/pubblicazioni/note-covid-19/2020/Nota-Covid-19.11.2020.pdf}.

\bibitem[Warren and Skillman(2020)]{Warren2020}
M.~S. Warren and S.~W. Skillman.
\newblock {Mobility Changes in Response to COVID-19}.
\newblock 2020.
\newblock URL \url{https://arxiv.org/abs/2003.14228}.

\bibitem[WHO(2004)]{WHO2004}
WHO.
\newblock {Guidelines on the use of vaccines and antivirals during influenza
  pandemics.}
\newblock Technical report, 2004.
\newblock URL
  \url{https://www.who.int/csr/resources/publications/influenza/11_29_01_A.pdf}.

\bibitem[{World Bank}(2021)]{WB2021}
{World Bank}.
\newblock {Global Economic Prospects, January 2021}, 2021.
\newblock URL
  \url{https://www.worldbank.org/en/publication/global-economic-prospects}.

\bibitem[Yoo and Managi(2020)]{yoo2020global}
S.~Yoo and S.~Managi.
\newblock Global mortality benefits of covid-19 action.
\newblock \emph{Technological Forecasting and Social Change}, 160:\penalty0
  120231, 2020.

\end{thebibliography}

\end{document}